\documentclass[11pt]{article}
\usepackage[T1]{fontenc}
\usepackage[utf8]{inputenc}
\usepackage[margin=1in]{geometry}
\usepackage{lmodern,microtype,amsmath,amssymb,amsthm,graphicx,booktabs,tabularx}
\usepackage[section]{placeins}
\usepackage[round,authoryear]{natbib}
\usepackage{xcolor}
\usepackage{marvosym}
\usepackage[colorlinks=true,linkcolor=blue!45!black,citecolor=blue!45!black,urlcolor=blue!45!black]{hyperref}
\newtheorem{proposition}{Proposition}
\newtheorem{hypothesis}{Hypothesis}
\newtheorem{corollary}{Corollary}
\newcommand{\E}{\mathbb{E}}
\newcommand{\R}{\mathbb{R}}
\newcommand{\sgn}{\operatorname{sgn}}

\newcommand{\TV}{\operatorname{TV}}

\title{Renormalized Cellular Intelligence: When Does Model Reduction Preserve Collective Capability?}
\author{Xuening Wu$^{1}$\thanks{\Letter~\href{mailto:xuening.wu@pfizer.com}  {\nolinkurl{xuening.wu@pfizer.com}}}, Yanlan Kang$^{2}$, Qianya Xu$^{3}$, Shan Yu$^{4}$, Haode Yang$^{1}$, Shenqin Yin$^{2}$\\[0.6em]
\small $^{1}$Pfizer Inc.\quad $^{2}$Fudan University\quad $^{3}$Cornell University\quad $^{4}$Independent researcher}
\date{Working manuscript, 24 October 2026}
\begin{document}
\maketitle
\begin{abstract}
As multi-agent systems grow in size and complexity, understanding which features of their interactions support collective capability becomes increasingly difficult. Can a simpler description preserve both task performance and the measured contribution of interactions between agents? Preserving collective dynamics alone is insufficient: reduction can alter decision-relevant information or give a non-communicating baseline access to pooled observations, distorting the measured advantage of interaction. We introduce Renormalized Cellular Intelligence (RCI), a framework for specifying and auditing capability preservation across scales. RCI treats the task, interacting dynamics, matched baseline, decision interfaces, and resource budget as a joint object of reduction. It specifies dynamical closure, decision compatibility, and preservation of the intended baseline comparison, and bounds the resulting error in relative capability gain. An exact analytical construction demonstrates task-score preservation, while a counterexample shows how closed reduced dynamics can erase the measured interaction gain. Using cellular systems as models of locally interacting agents, we evaluate three reduction candidates on five frozen learned models without retraining. The final candidate reduces the evolving state from eight to five dimensions per cell (37.5\%). Within-run uncertainty bounds support relative gain above a prespecified threshold for the fixed model collection, yet the candidate fails the closure and decision-preservation requirements for a complete RCI certificate. RCI provides explicit criteria for distinguishing positive interaction gains from faithful capability preservation; repeated renormalization and broader applicability remain open.
\end{abstract}

\section{Introduction}
Increasing the number of agents expands the possible interactions within a system, but does not reveal which of them support task performance. Some components may provide complementary information, while others may be redundant for a particular task. Understanding collective capability therefore requires descriptions that distinguish essential structure from dispensable detail. Such descriptions are useful only if simplification preserves the capability comparison they are intended to explain.

Preserving collective dynamics alone does not meet this challenge. A reduced model may discard information needed for task decisions or pool observations that were separate in a non-communicating baseline. It can therefore reproduce dynamics while distorting the measured advantage of interaction. This motivates the central question of Renormalized Cellular Intelligence (RCI): under what conditions does a simpler description preserve both task performance and the interaction-versus-baseline contrast?

This comparison requires a matched null: a declared system in which the relevant interactions are removed or restricted while observations, decisions, resources, and evaluation conditions are controlled. Merely disabling communication after reduction is insufficient if aggregation has already changed information access. The interacting system and its baseline must be reduced together.

\begin{figure}[htbp]
    \centering
    \includegraphics[width=0.8\linewidth]{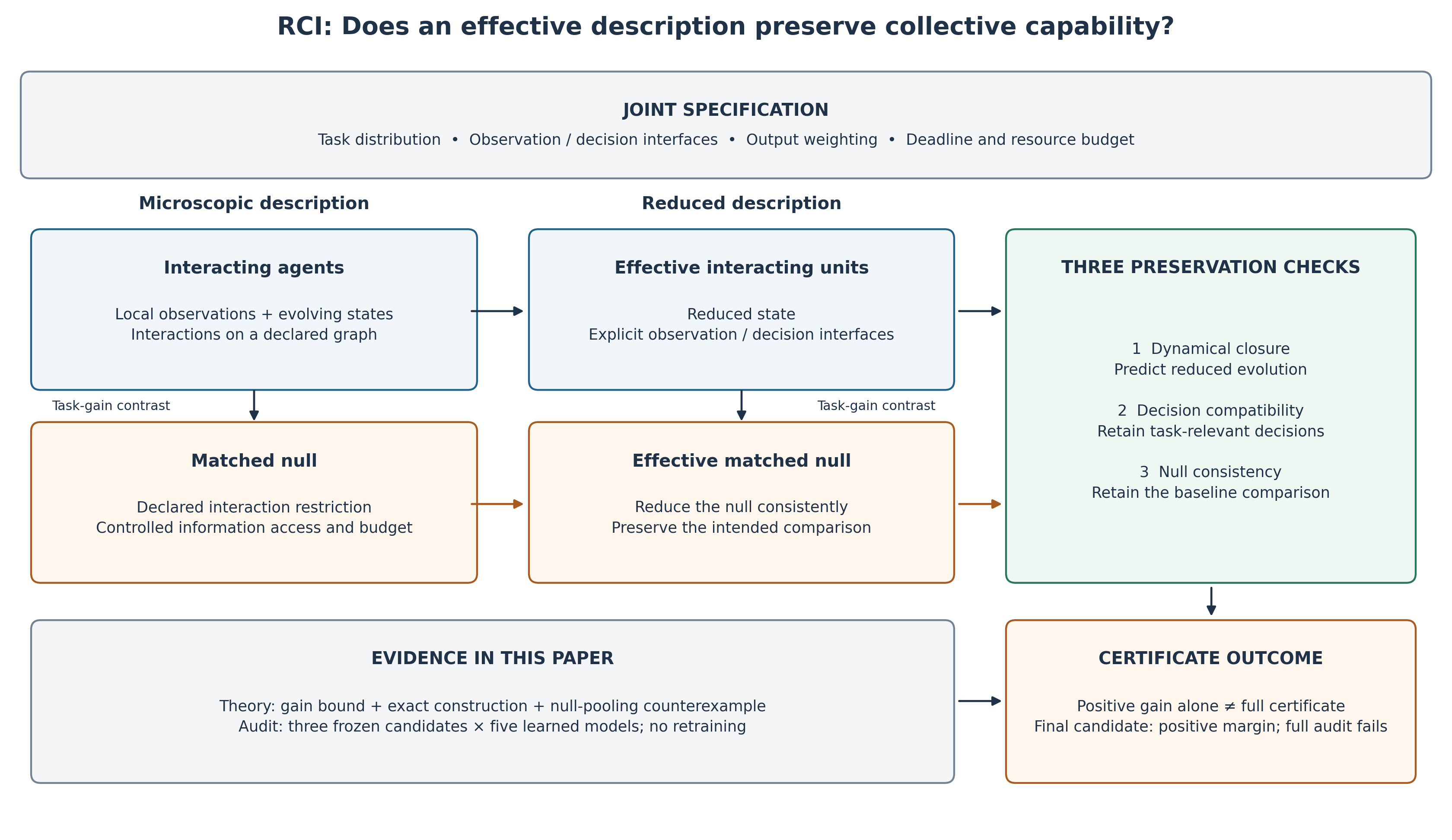}
    \caption{Overview of Renormalized Cellular Intelligence (RCI).
    The interacting system and its matched null are reduced jointly
    under a shared task and resource specification.
    Closure, decision compatibility, and null consistency
    assess whether the reduction preserves collective capability.
    A positive relative-gain margin alone does not constitute
    a complete RCI certificate.}
    \label{fig:rci-overview}
\end{figure}

We introduce Renormalized Cellular Intelligence (RCI), a framework for specifying and auditing capability preservation across scales. A cellular unit has local observations, internal state, and interactions on a declared graph; an effective cell represents a group through reduced state and an explicit interface. Cellular systems provide a controlled substrate for examining the broader problem of faithful descriptions of collective capability. Figure~\ref{fig:rci-overview} summarizes the RCI framework and the analytical and empirical evidence presented in this paper.

RCI treats the task distribution, interacting and null dynamics, decision interfaces, output weighting, deadline, and resource budget as a joint object of reduction. Its requirements are \emph{dynamical closure}, \emph{decision compatibility}, and \emph{null consistency}: the effective dynamics must satisfy the declared approximation, task decisions must remain compatible, and the intended baseline comparison must be preserved. RCI distinguishes relative interaction gain from practical capability onset and cross-scale composition. Positive gain alone establishes neither adequate absolute task quality nor a complete preservation certificate.

Our contributions are:
\begin{enumerate}
    \item A paired preservation specification and an expected-gain bound that connects approximation errors to a relative-gain threshold.
    \item An exact Gaussian four-to-two construction preserving closed dynamics and expected task gain, together with a counterexample showing how null information pooling destroys the contrast despite dynamical closure.
    \item An audit of three sequentially frozen reductions on five previously trained cellular models without retraining, retaining all failures. The final candidate reduces evolving-state dimension by 37.5\% and has a positive relative-gain margin under within-run uncertainty bounds, but fails the complete certificate because closure and decision-error requirements are unmet.
\end{enumerate}

RCI synthesizes established abstraction principles into an explicit interaction-versus-null preservation problem. Historical inference and inheritance studies provide supporting diagnostics, not additional successful reductions. Repeated renormalization and generalization to broader multi-agent architectures remain open.

\section{Related Work}
\paragraph{Cellular computation.}
Learned local rules support growth and regeneration \citep{mordvintsev2020}, with broader biological and computational connections reviewed by \citet{hartl2025}. \citet{dacre2025} study decentralized geometric sensing in a physical inductive array; \citet{etcheverry2026} demonstrate cellular reasoning on mazes, Sudoku, and ARC. These motivate task-level evaluation but are different systems from our Gaussian-localization benchmark. The comparison in Table~\ref{tab:recent-position} concerns declared objectives, not matched empirical performance or a comprehensive audit of their controls.

\paragraph{Coarse-graining and abstraction.}
Classical renormalization \citep{wilson1971}, cellular-automaton coarse descriptions \citep{israeli2006}, information bottleneck \citep{tishby2000}, and information-theoretic real-space RG \citep{kochjanusz2018,gordon2021} establish compression relative to relevant structure. Spectral graph reduction preserves structural and diffusion properties \citep{schmidt2024}. Approximate MDP homomorphisms bound performance loss \citep{taylor2008}; causal abstraction formalizes intervention consistency \citep{rubenstein2017}. Our closure and error arguments use these established principles. RCI applies them jointly to interacting and null dynamics, including admissible observations and decisions. Existing theories can represent such pairs; we do not claim a new general abstraction theorem or universal RG flow.

\paragraph{Information and adaptation.}
Gossip algorithms characterize distributed averaging \citep{boyd2006}; partial information decomposition distinguishes redundancy and synergy \citep{williams2010}; causal emergence compares effective-information descriptions \citep{hoel2013}. RCI instead measures an external task contrast. Its parity construction uses an engineered decoder, not a new synergy estimator, and task-score preservation does not create Shannon information. Inheritance and evolvability also have established precedents \citep{sasaki1999,bedau2003}. Adjacent cellular studies examine communication heterogeneity \citep{singh2026}, learnable novelty \citep{zhang2026}, and phenotype--genotype diversity \citep{jain2026}; our harvest and persistent-type pilot measures different observables.

\paragraph{Contribution boundary.}
RCI's contribution is the paired preservation specification, its constructive example and pooling obstruction, and an explicit frozen-model audit. The general framework comparisons are in Table~\ref{tab:position}; both positioning tables and the control hierarchy are retained in Appendix~\ref{app:position-tables}. They establish context, not superiority to existing reduction algorithms. None of our task scores measures general intelligence.

\section{RCI Framework and Methods}\label{sec:methods}
\subsection{Task-conditioned capability specification}
\subsubsection{Microscopic dynamics and the unit of evaluation}
Let $\mathcal G=(V,E)$ have $N$ units with states $x_i(t)\in\mathcal X_i$. A general local update is
\begin{equation}
 x_i(t+1)=F_{\theta_i}\!\left(x_i(t),\{m_{ji}(t):j\in\mathcal N_i\},o_i(t),\xi_i(t)\right),
 \qquad m_{ji}(t)=M_{ji}(x_j(t)),
\end{equation}
where $o_i$ is a local observation and $\xi_i$ is stochasticity. Asynchronous updates can be represented by including the update schedule in the state or transition kernel. State and parameter adaptation can be included in the local state, so that the Markov description contains the variables needed for learning as well as action. A task specifies a joint distribution $\mathcal D$ over environment variables, observations, and target $Y$, a decoder $d$, a deadline $T$, and a score $s(\widehat Y,Y)\in[0,1]$. A budget vector $\mathcal B$ records observations, state capacity, update opportunities, messages or bits, training cost, and calibration cost. Equalizing one component does not equalize the others.

For node-level distributed decisions, define
\begin{equation}
 Q(\mathcal S;\mathcal D,T,\mathcal B)=\E\left[\frac1N\sum_{i=1}^N s(d(x_i(T)),Y)\right],\qquad
 G_{\rm cap}=Q(\mathcal S)-Q(\mathcal S_0),\label{eq:gain}
\end{equation}
where $\mathcal S_0$ is a prespecified non-interaction null. The node average is a performance functional, not $N$ independent observations. An episode supplies one paired score difference. Replicate seeds form blocks when conditions share random numbers.

A matched null retains each unit's initial information, decoder, readout noise, state capacity, and update opportunities while disabling exchange. Communication removal with fixed learned weights tests dependence on communication; separately training a null tests achievable noncommunicating performance. These are different comparisons.

\subsubsection{What a matched null does and does not establish}
Table~\ref{tab:nulls} specifies the comparison hierarchy. A positive $G_{\rm cap}$ establishes benefit over the named null, not over every centralized algorithm. A centralized agent with all observations has a different information interface; it is useful as an information-access benchmark but must be separately budgeted for any efficiency claim. A serial implementation of the same decomposition tests whether parallel interaction, rather than decomposition alone, accounts for an advantage.

Agreement $A=\Pr(\widehat Y_i=\widehat Y_j)$, type count, persistence, and population size are diagnostics, not substitutes for a task score: collective agreement can repeat a wrong answer.

The Gaussian parity model, exact gain calculation, locality constraint, and descriptor-collision analysis used by the supporting studies are given in Appendix~\ref{app:inference-details}. These model-specific results illustrate the framework without defining a universal intelligence law.

\subsubsection{Performance gain, capability onset, and functional composition}
A positive expected $G_{\rm cap}$ means that interaction improves the specified score. This can be a small improvement on an existing function, or an improvement confined to a few nodes. It does not by itself establish acquisition of a qualitatively new function. For a fixed task distribution, deadline, null, and budget, select a practical margin $\delta>0$ before evaluation and define
\begin{equation}
 \mathcal C_\delta=\{\vartheta:G_{\rm cap}(\vartheta)>\delta\}.
 \label{eq:capregion}
\end{equation}
We call entry into this set \emph{practical gain onset}. If capability means solving a task to a required quality, also specify an absolute success level $q_\star$ and require $Q_1\geq q_\star$. The two criteria answer different questions: improvement over a baseline and adequacy for a task. A system can satisfy one without satisfying the other.

For an ordered coupling parameter, $\alpha_\delta=\inf\{\alpha:G_{\rm cap}(\alpha)>\delta\}$ describes the first crossing when the set is nonempty. It need not be the only boundary: the region can be disconnected and gain need not be monotone. If broadly distributed capability is intended, an additional coverage condition such as $\Pr_{i\sim\omega}(Q_{1,i}\geq q_\star)\geq\kappa$ can be specified in advance. The scalar mean score in this paper does not enforce coverage on its own.

Cross-scale functional composition additionally requires an explicit relation between tasks or interfaces at different levels; a positive gain or threshold crossing does not establish it.

Onset requires uncertainty control. The archived $\delta=.02$ precision/recall measurements are descriptive; they do not supply simultaneous boundary guarantees. Corollary~\ref{cor:margin} controls approximation error, not statistical uncertainty or thermodynamic criticality.

\begin{hypothesis}[Task-conditioned micro-to-macro prediction]
Within a declared family of tasks and local dynamics, a frozen response descriptor predicts held-out capability gain within a prespecified error tolerance across information layouts, deadlines, and graph sizes.
\end{hypothesis}
Probe 2 supports this hypothesis for its finite Gaussian test family. Probe 3 rejects the stronger version that omits the latent/observation distribution from the family specification.

\begin{hypothesis}[Robust practical onset]
A descriptor-based predictor locates a prespecified positive-gain region under new conditions, with acceptable false-positive rate and uncertainty-calibrated boundaries, beyond easy zero-communication cases.
\end{hypothesis}
This remains stronger than the archived precision and recall measurements. A thermodynamic critical point would require additional finite-size and limiting arguments; no such claim follows from these hypotheses.

\subsection{Task-preserving reduction}\label{sec:rg}
\subsubsection{The experimental specification as the object of reduction}
A low-dimensional regression on experimental conditions does not by itself construct a smaller dynamical system. For a state reduction, let $X_t$ be the joint microscopic state, $\mathsf P_a$ its Markov kernel in arm $a\in\{1,0\}$, and $\pi_b:X\mapsto Z$ a block map. Arm 1 denotes interaction and arm 0 its designated null. The transformation must act on the complete experimental specification,
\begin{equation}
 \mathcal E=(\mathcal D,\mathsf P_1,\mathsf P_0,d_1,d_0,\omega,T,\mathcal B)
 \quad\longmapsto\quad \overline{\mathcal E}_b,
 \label{eq:experimenttuple}
\end{equation}
where $\omega$ weights output units. The microscopic default is $\omega_i=1/N$. The effective specification includes transformed state distributions, dynamics, admissible observations, decoders, weights, time, and budget. A macro representation is free to use fewer variables, but this does not imply that an effective cell has the information or resource cost of one original cell.

Exact closure over $\tau_b$ microsteps requires, for every measurable macrostate set $D$ and both arms,
\begin{equation}
 \mathsf P_a^{\tau_b}(x,\pi_b^{-1}(D))
 =\overline{\mathsf P}_{a,b}(\pi_b(x),D)
 \quad\text{for every }x.\label{eq:closure}
\end{equation}
This is a strong lumpability-type condition. If the condition fails, a memoryless macrostate is not exact. An approximate description must bound the error or retain additional boundary states or memory. For deterministic linear dynamics and a block-averaging matrix $L$, the corresponding intertwining condition is
\begin{equation}
 LP_a^{\tau_b}=\overline P_{a,b}L.\label{eq:intertwine}
\end{equation}
Even exact intertwining need not preserve task performance: a decoder can use distinctions or observations that the projection changes.

\subsubsection{Decision preservation and the matched null}
For Bayes-optimal decisions, a sufficient state satisfies
\begin{equation}
 \Pr(Y\mid X_T)=\Pr(Y\mid Z_T),\qquad Z_T=\pi_b(X_T),\label{eq:suff}
\end{equation}
under the relevant arm and task distribution. This preserves optimal decisions for bounded losses. It need not preserve a particular implemented decoder. For such a decoder, define the noise-averaged microscopic score $h_a(X_T,Y)$ and effective score $\overline h_a(Z_T,Y)$. The directly relevant requirement is equality, or a controlled difference, of their expectations in each arm. Randomized readouts can be included in the state or integrated into $h_a$; the choice must be consistent. The default node average becomes a weighted block average, with $\overline\omega_B=|B|/N$, unless a different output experiment is explicitly intended.

The data-processing inequality gives $I(Y;Z_T)\leq I(Y;X_T)$ for deterministic $\pi_b$. Better performance by a restricted learner after compression concerns its decision class or inductive bias, rather than creation of information. In particular, global sufficiency of a macrostate does not imply preservation of each unit's local information access.

Let $\mathcal N$ remove the designated exchange channel and $\overline{\mathcal N}_b$ be its effective counterpart. The operational requirement is
\begin{equation}
 \mathcal R_b\bigl(\mathcal N(\mathcal E)\bigr)
 \simeq \overline{\mathcal N}_b\bigl(\mathcal R_b(\mathcal E)\bigr).\label{eq:nullcommute}
\end{equation}
Here equivalence includes admissible observations and decision scores as well as transition laws. A projected null that gives one effective cell simultaneous access to previously separated complementary observations may change the null task. The counterexample in Section~\ref{sec:counterexample} shows why a commuting diagram for state dynamics alone cannot establish this operational equivalence.

\paragraph{Two declared levels of null preservation.}
Task-score null equivalence preserves the expected score for the specified task, decoder, deadline, and weights. The parity example proves this guarantee only. A port-isolated null additionally restricts every labeled private state to its own state and observations when communication is removed. The learned-model candidates impose this restriction and separately test dynamics and decisions. Neither level implies the other; every reduction must declare which guarantee it supplies.

\subsubsection{Approximation errors and practical onset}
\begin{proposition}[A sufficient bound for capability preservation]\label{prop:bound}
For each arm $a\in\{1,0\}$, suppose the joint endpoint laws of $(Y,Z_T)$ under the projected micro process and the effective macro process differ in total variation by at most $\epsilon_a$. Under the projected micro law, suppose the difference between the expected original score and a common effective score in $[0,1]$ is at most $\eta_a$. Then
\begin{equation}
 |G_{\rm micro}-G_{\rm macro}|
 \leq \epsilon_1+\epsilon_0+\eta_1+\eta_0=:E_b.\label{eq:gainbound}
\end{equation}
If initial joint-law error is at most $a_a$ and the task-conditioned one-step coarse-kernel error is uniformly at most $e_a$ over $m$ macrosteps, then $\epsilon_a\leq\min(1,a_a+me_a)$.
\end{proposition}
\begin{proof}
A common $[0,1]$-valued score changes in expectation by at most the total-variation distance, using $\TV(\mu,\nu)=\sup_D|\mu(D)-\nu(D)|$. Add the score-compatibility error in each arm and apply the triangle inequality to the interaction--null difference. The multi-step statement follows by contraction under a Markov kernel and induction. It requires comparison on the joint state retaining $Y$, with compatible conditional kernels; marginal closure without the target is insufficient.
\end{proof}

\begin{corollary}[Preservation of a relative-gain classification]\label{cor:margin}
Fix $\delta>0$. If $G_{\rm micro}>\delta+E_b$, then $G_{\rm macro}>\delta$. If $G_{\rm micro}<\delta-E_b$, then $G_{\rm macro}<\delta$. Thus the classifications agree whenever $|G_{\rm micro}-\delta|>E_b$.
\end{corollary}
\begin{proof}
Apply Eq.~\eqref{eq:gainbound} to the upper or lower endpoint of the interval around $G_{\rm micro}$.
\end{proof}
This is a deterministic approximation guarantee for expected gains. Statistical uncertainty in estimated gains or errors must be added separately; a fitted predictor's RMSE is not $E_b$. The interval around $\delta$ identifies an unresolved margin, rather than a critical point.

For deterministic continuous dynamics, total variation can be uninformative despite small Euclidean errors. Appendix~\ref{app:continuous-errors} gives the task-conditioned Wasserstein alternative and the parity decoder's noise-smoothed Lipschitz bound. The learned-model experiment uses the directly coupled localization-distance bound instead.

\subsection{Analytical constructions}
The paired preservation criteria yield an exact example and a contrasting null-pooling obstruction.

\subsubsection{An exact four-to-two analytical construction}\label{sec:analytic}
The following is a symbolic example, not an additional simulation or a fitted result. It illustrates exact closure and task-score preservation on a four-node graph, at the task-score null-equivalence level just defined. Here $Y=\sgn(A)\sgn(B)$, and an endpoint predicts $\widehat Y=\sgn(z_A+\eta_A)\sgn(z_B+\eta_B)$ with independent $\eta_A,\eta_B\sim\mathcal N(0,\sigma^2)$. Appendix~\ref{sec:gaussian} derives the general Gaussian score; this example uses a deliberately chosen stochastic matrix.

\paragraph{Microscopic states and the map.}
Nodes 1 and 2 observe $A+\varepsilon_1,A+\varepsilon_2$, and nodes 3 and 4 observe $B+\varepsilon_3,B+\varepsilon_4$, with independent unit Gaussian sensor noise and independent standard normal $A,B$. Initial channel vectors are
\begin{align}
 x^{(A)}_0&=(A+\varepsilon_1,A+\varepsilon_2,0,0)^{\mathsf T},\\
 x^{(B)}_0&=(0,0,B+\varepsilon_3,B+\varepsilon_4)^{\mathsf T}.
\end{align}
Group equal-role nodes into two blocks and apply the same map in both channels:
\begin{equation}
 L=\frac12\begin{pmatrix}1&1&0&0\\0&0&1&1\end{pmatrix},\qquad
 U=\begin{pmatrix}1&0\\1&0\\0&1\\0&1\end{pmatrix},\qquad LU=I_2.
 \label{eq:LU}
\end{equation}
The effective state is $z^{(r)}_t=Lx^{(r)}_t$. The discarded variables are the two within-block differences in each channel. Define $J_2=\tfrac12\boldsymbol 1\boldsymbol 1^{\mathsf T}$ and, for $0<\gamma<1$,
\begin{align}
 \overline P_{\gamma}&=\begin{pmatrix}1-\gamma&\gamma\\\gamma&1-\gamma\end{pmatrix},\\
 P_{\gamma}&=U\overline P_{\gamma}L
 =\begin{pmatrix}(1-\gamma)J_2&\gamma J_2\\\gamma J_2&(1-\gamma)J_2\end{pmatrix}.
 \label{eq:fourmatrix}
\end{align}
Weights are local to the declared four-node graph; this construction makes no lattice-limit claim. Take one microscopic round to be one effective round. Multiplication gives
\begin{equation}
 LP_{\gamma}=\overline P_{\gamma}L,\qquad P_{\gamma}U=U\overline P_{\gamma},\qquad
 P_{\gamma}^t=U\overline P_{\gamma}^{\,t}L\quad(t\geq1).
 \label{eq:exactreduction}
\end{equation}
Thus the two effective units have exact closed dynamics. After one interaction round the microscopic state is block-constant and satisfies $x^{(r)}_t=Uz^{(r)}_t$. For clarity, subscripts 0 and 1 continue to denote experimental arms, whereas $\gamma$ denotes the mixing fraction. For the original self-copy null $P_0=I_4$, the effective null is $\overline P_0=I_2$, since $LI_4=I_2L$. The same map therefore closes both arms, even though the null does not erase within-block differences.

\paragraph{Sensor covariance, readout, and score.}
Let $\overline\varepsilon_A=(\varepsilon_1+\varepsilon_2)/2$ and $\overline\varepsilon_B=(\varepsilon_3+\varepsilon_4)/2$. Their variances are $1/2$ and they are independent. For the first interaction round,
\begin{align}
 z^{(A)}_1&=((1-\gamma)(A+\overline\varepsilon_A),\ \gamma(A+\overline\varepsilon_A))^{\mathsf T},\\
 z^{(B)}_1&=(\gamma(B+\overline\varepsilon_B),\ (1-\gamma)(B+\overline\varepsilon_B))^{\mathsf T}.
\end{align}
The effective distribution must retain the shared sensor noise across its two endpoints; for example,
\begin{equation}
 \operatorname{Cov}(z^{(A)}_1\mid A)=\frac12
 \begin{pmatrix}(1-\gamma)^2&\gamma(1-\gamma)\\\gamma(1-\gamma)&\gamma^2\end{pmatrix}.
\end{equation}
Each endpoint decoder receives independent Gaussian readout noise of variance $\sigma^2$ per channel, exactly as in the microscopic output experiment. Projection is applied \emph{before} readout; averaging two microscopic readout noises would instead change the experiment. Effective output weights are $1/2$ and $1/2$, since each block contains two original nodes.

Block-constant interacting states give identical noise-averaged scores to their effective cells. In the null, each block has only one role; the missing channel is independent symmetric readout noise, so both scores are $1/2$. Same-role pooling improves sensor precision without revealing the missing latent sign. This proves task-specific expected-score compatibility, not preservation of all null states, other tasks, or the original private-port partition.

\begin{proposition}[Exact expected-gain preservation in the four-to-two construction]\label{prop:example}
For the task, dynamics, output weighting, and readout experiment just specified, the microscopic and effective expected gains after one round are equal:
\begin{align}
 G_{\rm micro}=G_{\rm macro}&=\frac12 r(u_+)r(u_-),\label{eq:examplegain}\\
 u_+&=\frac{1-\gamma}{\sqrt{(1-\gamma)^2/2+\sigma^2}},\qquad
 u_-=\frac{\gamma}{\sqrt{\gamma^2/2+\sigma^2}}.
\end{align}
where $r(u)=2\arctan(u)/\pi$. The gain is positive for $0<\gamma<1$ and $\sigma>0$.
\end{proposition}
\begin{proof}
The first effective cell's two channel ratios are $u_+,u_-$, and the second cell's ratios are exchanged. Proposition~\ref{prop:gain} gives the common gain formula. Block-constant interacting states and the original node weights give the same microscopic average, while both null scores equal $1/2$. Positivity follows because both ratios are positive.
\end{proof}
At $\gamma=0$ and $\gamma=1$, each endpoint still receives only one role and gain is zero. A positive interior region is therefore possible without monotonicity in coupling or a phase transition. The construction preserves expected task capability, not the joint distribution of every node's answers or all agreement statistics. It also does not show that an autonomous effective cell can reproduce the system at the cost of a single original sensor: each effective cell retains the provenance of two sensors, and the budget ledger retains the four-sensor microscopic experiment. These are changes in representation, not measured savings in work or energy.

\subsubsection{Closed coarse dynamics can destroy the capability contrast}\label{sec:counterexample}
Consider two nodes, one observing $A+\varepsilon_A$ in channel $A$ and one observing $B+\varepsilon_B$ in channel $B$. Set $P_1=J_2$, $P_0=I_2$, and project both nodes to a single effective cell with $L_*=(1/2,1/2)$. Then
\begin{equation}
 L_*P_1=L_*=L_*P_0.
\end{equation}
Both projected dynamics are exactly the one-cell identity. Nevertheless, their operational decision experiments are not the same as the originals. In the microscopic interacting arm, each node receives $(A+\varepsilon_A)/2$ and $(B+\varepsilon_B)/2$ before readout. With the same per-channel readout variance $\sigma^2$,
\begin{equation}
 Q_1=\frac12+\frac12r(u_*)^2,\qquad
 Q_0=\frac12,\qquad
 u_*=(1+4\sigma^2)^{-1/2}.
\end{equation}
The microscopic gain is strictly positive. A naive macro decoder with access to both components of $L_*X_0$ already sees both complementary observations in the effective null. It has the same score $Q_1$ in both effective arms, yielding
\begin{equation}
 \overline G_{\rm naive}=0\neq G_{\rm micro}=\frac12r(u_*)^2.
 \label{eq:nullfailure}
\end{equation}
The dynamical projections commute perfectly. The failure is null-score compatibility: aggregation silently pools evidence before the supposed removal of interaction. In Proposition~\ref{prop:bound}, the dynamical errors can be zero, but the null decision error is nonzero and accounts for the lost contrast.

A faithful reduction must retain restricted null ports, verify a task-equivalent randomized null decoder, or reject this aggregation. A decoder matching one expected score is weaker than preservation across a task family: same-role pooling cannot reveal the missing sign, whereas mixed-role pooling can.

\subsection{Effective units, composition, and the open RCI hypothesis}
An effective cell is task-relative. Eliminating variables may require boundary state, memory, or longer-range interactions; these belong in the effective dynamics.

Repeated reduction requires compatible interfaces, nulls, budgets, and propagated errors, with $\mathcal R_{b_1}\mathcal R_{b_2}\simeq\mathcal R_{b_1b_2}$ under specified coordinate/time conventions. One exact map does not establish this composition law or an RG fixed point.

\begin{hypothesis}[Task-preserving cross-scale description]\label{hyp:rg}
For a declared cellular model and task family, nontrivial successive block maps reduce state dimension while keeping closure, decision, and null-consistency errors small enough to preserve practical capability classification at multiple scales.
\end{hypothesis}
A proposed map can fail because two states in the same block class have different task-conditioned futures, because a decoder loses relevant information, or because aggregation changes the null's information access. These are different failure modes with different repairs. They should be checked before interpreting a regression collapse or fitting critical exponents.

An adaptive extension must distinguish inherited solutions from inherited capacity to learn. Harvest and persistent types do not resolve that question.

Section~\ref{sec:endtoend} tests one reduction level. Neither graph-size prediction transfer nor the exact construction establishes repeated multiscale closure or critical scaling.

\section{Experiments and Results}\label{sec:endtoend}
\subsection{Task, frozen systems, and the paired decision experiment}
We evaluated explicit reductions of five already-trained cellular models, retaining seeds 1000--1004 without selecting a favorable model or updating neural parameters. Each system has a $32\times32$ open-boundary lattice with eight state channels per site. Let $u_i\in[0,1]^2$ be a site's coordinates and $Y\sim\mathrm{Uniform}([0,1]^2)$ a localization target. A static observation is
\begin{equation}
 O_i=\left(M_i\{\exp[-\|u_i-Y\|_2^2/(2s^2)]+\nu_i\},M_i,u_i\right),
 \quad s=0.1875,\quad M_i\sim\mathrm{Bernoulli}(0.2),\quad \nu_i\sim\mathcal N(0,0.1^2).
\end{equation}
Masks and sensor noise are sampled independently for each repeat; all compared paths use the same observations. Let $A$ be the degree-normalized four-neighbor mean. For $X_0=0$,
\begin{equation}
 F_\lambda(X;O)=\tfrac12 X+\tfrac12\phi_\theta(((1-\lambda)I+\lambda A)X,O),\qquad
 \phi_\theta(c,O)=\tanh\!\left(W_2\tanh(W_1[c,O]+b_1)+b_2\right).
 \label{eq:nca}
\end{equation}
The rule has widths $12\to32\to8$ and 680 trainable scalars, all frozen during the present evaluations. Historical training used supervised backpropagation through the 32-step unroll; local inference does not imply a locally learned rule. Appendix~\ref{app:training-origin} reports the fingerprint-verified training settings, checkpoint choices, and unavailable details. Interaction uses $\lambda=0.5$; the null switches the same weights to $\lambda=0$. This is a test of a trained rule's dependence on communication, not a comparison against the strongest separately trained noncommunicating model.

The implemented decision is the \emph{same global readout} in both arms,
\begin{equation}
 q(X)=\frac1{1024}\sum_i\frac{X_{i,1:2}+1}{2},\qquad
 \ell(q,Y)=\frac{\|q-Y\|_2}{\sqrt2},\qquad Q=1-\E\ell.
 \label{eq:nca-readout}
\end{equation}
This extends the node-level functional in Eq.~\eqref{eq:gain} to a specified collective decoder. Global averaging is allowed at the terminal decision in both arms; it does not authorize additional hidden-state communication during a null rollout. All paths run for 32 updates. The target of inference is the expected performance of the \emph{fixed, equally weighted five-model collection}, not the population of possible training seeds.

\subsection{Three sequential candidates and their budgets}
Each candidate was frozen before its own 1,024-target evaluation with four observation repeats and all five fixed models. Candidate development was sequential and informed by preceding failures, rather than three independent replications of one untouched hypothesis. Complete historical training inputs are unavailable; distinct streams and checks against accessible targets do not certify exclusion from the full historical training stream.

The block-mean candidate (v1) averages each $2\times2$ block to eight channels and broadcasts it for the next update, creating within-block hidden-state information paths even in the null. The four-dimensional candidate (v2) instead keeps four labeled private ports, retaining two output coordinates and two leading hidden-state singular directions per port. The five-dimensional candidate (v3) adds one private residual direction derived only from old balanced fit trajectories. It evolves autonomously; no microscopic teacher state supplies the fifth coordinate during rollout. Projection definitions and fit provenance are in Appendix~\ref{app:frozen-methods}.

All candidates retain 1,024 raw observation ports and the original sitewise MLP evaluations. The private-port candidates are compound units with internal interfaces, not replacement of four sensors by one isolated sensor. Table~\ref{tab:budget-ledger} reports evolving-state capacity only.

\begin{table}[htbp]\centering\small
\caption{Resource ledger for the executed candidates. Counts refer to evolving state scalars, not full implementation memory or measured computational cost. All rows retain the same 32-step deadline and 680-parameter microscopic rule.}\label{tab:budget-ledger}
\begin{tabular}{lrrrr}\toprule
System & State scalars & State factor & Raw ports & Site MLPs/step\\\midrule
Microscopic & 8,192 & $1$ & 1,024 & 1,024\\
v1 & 2,048 & $4$ & 1,024 & 1,024\\
v2 & 4,096 & $2$ & 1,024 & 1,024\\
v3 & 5,120 & $1.6$ & 1,024 & 1,024\\
\bottomrule\end{tabular}
\end{table}
The state factor in Table~\ref{tab:budget-ledger} is $8192/d$, where $d$ counts evolving scalars. It excludes fixed projection matrices, observation storage, temporary lifted states, and audit trajectories. Every reduced step still lifts to the original sites, evaluates the original local MLPs, and projects back. These extra operations prevent inferring an arithmetic-speed benefit from the state factor. Deployment FLOPs, peak memory, wall-clock latency, and energy were not measured in a matched benchmark. Execution-archive elapsed times, which also include microscopic reference paths and certificate diagnostics, cannot substitute for such a benchmark.

\subsection{Evaluation criteria and uncertainty}
Let $X_t^a$ and $Z_t^a$ be the autonomous microscopic and reduced paths, with arm $a=1$ for interaction and $a=0$ for the null. For each trajectory, $C_a=\max_{t<32}\max_j\|r_{a,j}(t)\|_2/(2\sqrt8)$ is the normalized maximum projected one-step residual (Eq.~\eqref{eq:empclosure}); $j$ labels blocks in v1 and private ports in v2/v3. Teacher states evaluate $r_a$ and never replace $Z_t^a$. The endpoint distance is $D_a=\|q(X_{32}^a)-\overline q(Z_{32}^a)\|_2/\sqrt2$, using the microscopic and reduced readouts. These distances bound gain error:
\begin{equation}
 |G_{\rm micro}-G_{\rm macro}|\leq\E D_1+\E D_0.
\end{equation}
The bound follows from the reverse triangle inequality for the normalized localization loss. Agreement between two null implementations and mathematical private-port isolation are separate from approximation of microscopic null dynamics or decisions. Neither implementation check removes the need to measure $C_0$ and $D_0$.

\paragraph{Confidence bounds and gates.}
Each independent target contributes the mean over four repeats and the five fixed models. Six one-sided empirical Bernstein bounds use $n=1024$ and $\beta=0.05/6$, giving within-run simultaneous 95\% coverage conditional on that collection and projection. Supplementary per-model bounds use $0.05/30$. Neither statement is simultaneous across the adaptively developed three-candidate sequence. The full formula and bounded-statistic ranges are retained in Appendix~\ref{app:frozen-methods}.

Here $G_L$ is a lower confidence bound on the expected microscopic gain $G_{\rm micro}$, and subscript $U$ denotes an upper confidence bound on the corresponding expected error or loss. Thus $C_{a,U}$ and $D_{a,U}$ bound $\E C_a$ and $\E D_a$. The frozen tolerances are
\begin{equation}
 C_{1,U}\leq0.05,\quad C_{0,U}\leq0.05,\quad
 B_U:=D_{1,U}+D_{0,U}\leq0.05,\quad
 M:=G_L-0.02-B_U>0,\quad \ell_{\rm macro,1,U}\leq0.15.
 \label{eq:allgates}
\end{equation}
These numerical tolerances are \emph{prespecified methodological audit choices for this synthetic task}, rather than universal constants of RCI or application-calibrated requirements. The relative-gain threshold $0.02$ declares the minimum gain to be certified; the two closure tolerances $0.05$ are expressed in the declared original-channel normalization; and the decision budget $0.05$ bounds the sum of the two normalized endpoint discrepancies. The statistical absolute-quality gate $\ell_{\rm macro,1,U}\leq0.15$ is equivalently $Q_{\rm macro,1,L}:=1-\ell_{\rm macro,1,U}\geq0.85$; the point estimate alone does not meet this confidence requirement. Their numerical roles are explicit, but no external application utility or safety requirement was used to validate these cutoffs. A different application must justify and freeze its own tolerances before evaluation. We retain all original cutoffs after seeing the failures; a post-hoc relaxation would define a different contract and cannot change this certificate.

The complete empirical RCI certificate requires every gate and implementation check. A positive $M$ certifies only the relative-gain threshold. Closure is additional to the endpoint-distance bound and is distribution-relative over this finite horizon, not uniform over all states. Vacuous global Lipschitz bounds remain reported without outcome-based fitting.

\subsection{Observed results and failure retention}
Table~\ref{tab:reduction-summary} reports every frozen candidate. No neural parameters were updated, no target or model was removed, and no six- or seven-dimensional candidate was launched. Numerical entries are recomputed from saved observation-level arrays.
\begin{table}[htbp]\centering\small
\caption{Frozen paired-reduction certificates. Entries are within-run simultaneous one-sided bounds, except the derived margin. All candidates retain all five models. Different interfaces and formal target streams prevent interpreting closure differences as matched improvement factors.}\label{tab:reduction-summary}
\begin{tabular}{lrrrrrl}\toprule
Candidate & $C_{1,U}$ & $C_{0,U}$ & $B_U$ & $M$ & $\ell_{1,U}$ & Overall\\\midrule
Block mean & 0.1362 & 0.1867 & 0.1371 & 0.0659 & 0.1198 & Fail\\
Four private dims & 0.1180 & 0.1153 & 0.3064 & -0.1059 & 0.2406 & Fail\\
Five private dims & 0.0720 & 0.0753 & 0.1955 & 0.0069 & 0.1666 & Fail\\
\bottomrule\end{tabular}
\end{table}
The final candidate did not pass the full RCI certificate. Its mean closure errors are 0.059428 (interaction) and 0.062696 (null). Mean endpoint distances are 0.085890 and 0.082877, with upper bounds 0.099983 and 0.095503. The gain lower bound is 0.222423, yielding $M=0.006937$. The mean macro gain is 0.217890; the sign of the conservative margin must not be confused with the sign of this observed gain. The final run retains all 5,120 target--model cases, including 4,649 cases failing at least one descriptive pointwise condition. These case labels are not per-target confidence certificates. All 20,480 observation--model records and their trajectories are retained in the execution archive. A positive margin certifies the prespecified relative-gain threshold alone: the same within-run bounds give $G_{\rm macro}\geq 0.026937>0.02$. This does not satisfy the additional frozen closure, decision-budget, or absolute-quality requirements. In particular, the point estimate $\widehat Q_{\rm macro,1}=1-0.146925=0.853075$ is above 0.85, but its lower confidence bound is only $Q_{\rm macro,1,L}=1-0.166611=0.833389<0.85$. Absolute task adequacy is therefore not certified; this is not a proof that its true value is below 0.85. Macro-null implementation agreement was 4.77e-07, projection-readout agreement was 2.38e-07, and the preflight three-step cross-port null perturbation effect was 0; all are within the frozen $10^{-6}$ tolerance. No checkpoint changed. These implementation results do not override the scientific gates.

For the final five-dimensional candidate, Table~\ref{tab:five-models} retains every model. The primary claim concerns the fixed collection; these supplementary rows cannot replace it with its best member. For example, seed 1003 has means below the closure tolerances and a positive margin, yet its corrected upper bounds are $C_{1,U}=0.062431$, $C_{0,U}=0.065713$, and $B_U=0.072964$; its complete certificate is also Fail.
\begin{table}[htbp]\centering\small
\caption{Five-dimensional candidate: per-model means for closure, decision distance, and macro gain. $M_s$ uses supplementary $0.05/30$ bounds; the final column evaluates every frozen gate using those bounds, not the displayed means. No individual model passes the complete certificate.}\label{tab:five-models}
\begin{tabular}{lrrrrrrl}\toprule
Seed & $C_1$ & $C_0$ & $D_1$ & $D_0$ & $G_{\rm macro}$ & $M_s$ & Full cert.\\\midrule
1000 & 0.0469 & 0.0558 & 0.0284 & 0.0195 & 0.2421 & 0.0944 & Fail\\
1001 & 0.0779 & 0.0818 & 0.1622 & 0.1527 & 0.1278 & -0.2309 & Fail\\
1002 & 0.0540 & 0.0559 & 0.1074 & 0.0667 & 0.2219 & -0.0346 & Fail\\
1003 & 0.0460 & 0.0494 & 0.0366 & 0.0028 & 0.3756 & 0.2375 & Fail\\
1004 & 0.0725 & 0.0706 & 0.0949 & 0.1727 & 0.1220 & -0.1466 & Fail\\
\bottomrule\end{tabular}
\end{table}

\begin{figure}[htbp]\centering
\IfFileExists{figures/rci_v3_task_trajectories.png}{\includegraphics[width=0.7\linewidth]{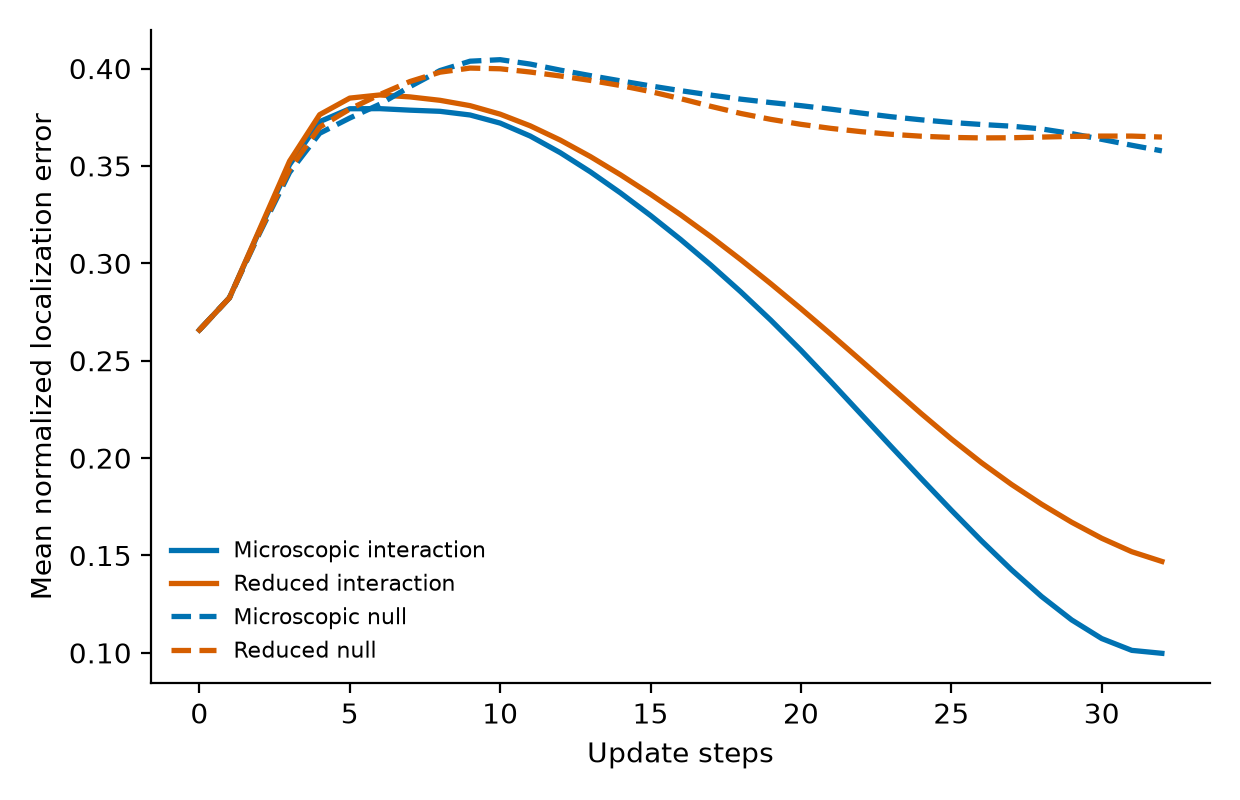}}{\fbox{\parbox{.88\linewidth}{\centering Figure supplied with the source package: \texttt{figures/rci\_v3\_task\_trajectories.png}. Numerical results are included in the tables and text.}}}
\caption{Final frozen candidate: mean localization loss over 32 updates, averaging all targets, repeats, and five fixed models. Both reduced branches roll autonomously; dashed curves are the matched same-weight null. Lower loss is better.}\label{fig:v3-trajectories}
\end{figure}
\begin{figure}[htbp]\centering
\IfFileExists{figures/rci_v3_error_budgets.png}{\includegraphics[width=0.8\linewidth,height=.77\textheight,keepaspectratio]{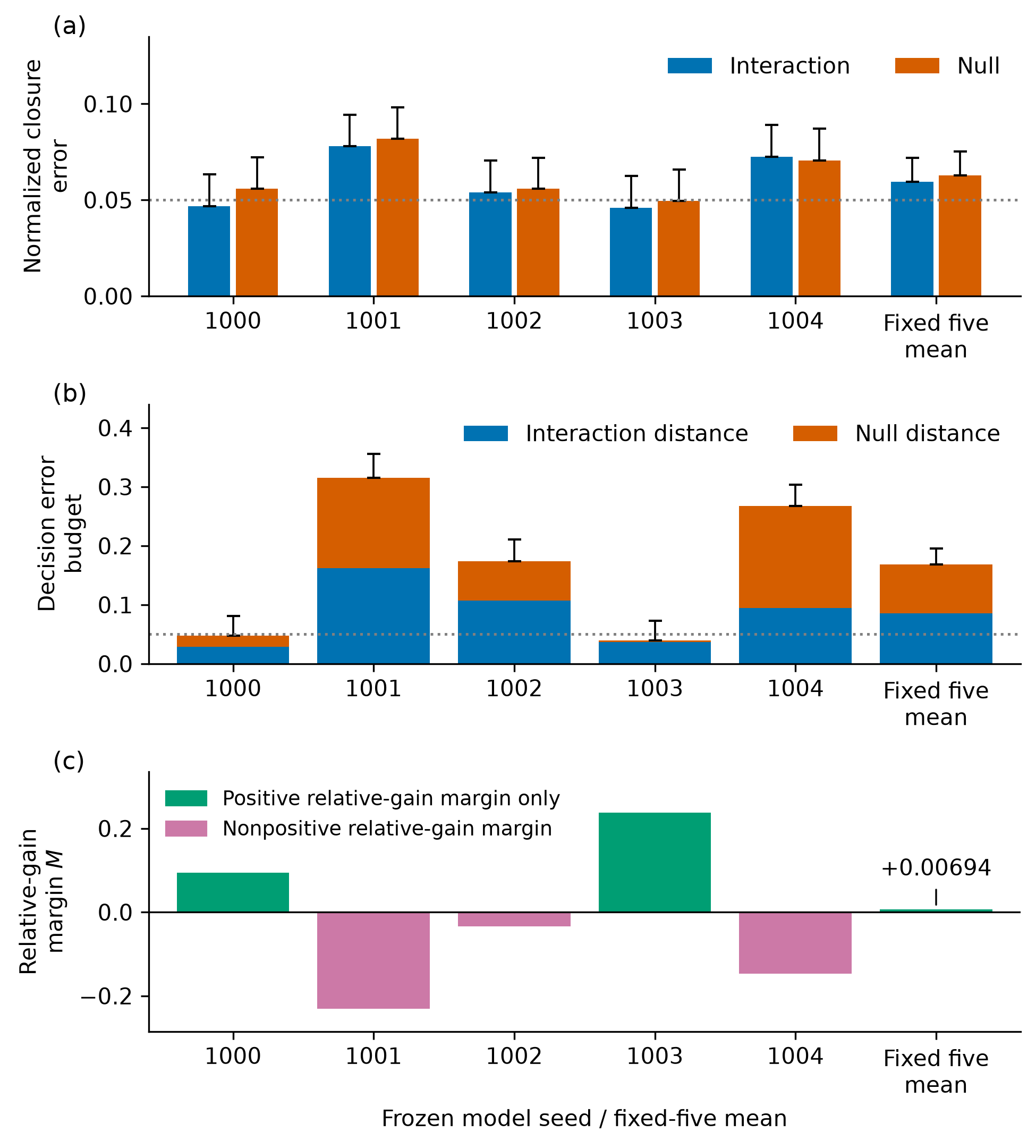}}{\fbox{\parbox{.88\linewidth}{\centering Figure supplied with the source package: \texttt{figures/rci\_v3\_error\_budgets.png}. Numerical results are included in the tables and text.}}}
\caption{Frozen five-dimensional candidate, shown at enlarged size. (a) Bars are mean closure errors; black whiskers end at the corresponding upper confidence bounds. (b) The stacked bars are mean interaction and null decision distances; whiskers end at their summed upper bound. Dotted lines mark the 0.05 tolerances. (c) $M$ is the conservative \emph{relative-gain} margin, not an overall certificate score: green means only $M>0$. All five individual models and the fixed-five collection fail the complete RCI certificate. Individual bounds use $0.05/30$; collection bounds use $0.05/6$. ``Fixed five mean'' averages the five models' metrics, not their predictions into a new decoder.}\label{fig:v3-errors}
\end{figure}

Appendix~\ref{app:horizon} diagnoses intermediate times from saved v3 paths. These post-hoc summaries supply no new certificate; $T=32$ remains the frozen deadline.

\paragraph{Failure interpretation.}
Readout identity and private-port isolation do not guarantee preservation of future decisions: discarded hidden components alter nonlinear feedback. The residual and state-error recurrence are given in Appendix~\ref{app:frozen-methods}. State reconstruction, interface correctness, closure, and task preservation are distinct checks. These failures do not disprove the existence of an adequate reduction; they also do not establish repeated multiscale composition.

\subsection{Supporting empirical case studies}\label{sec:experiments}

These archived studies complement the reduction results by identifying
distinctions an RCI description must retain.
Their complete designs, results, controls, and limitations remain in
Appendix~\ref{app:case-details}.

The inference probes show that propagation summaries alone can miss
where complementary observations are located and whether they remain
aligned with the task. Agreement can persist when useful gain
disappears. Task-aligned co-reachability improves prediction within
the tested Gaussian family, but moment-matched distribution changes
defeat the frozen predictor even with additional calibration.
These findings support task-aware descriptors while exposing their
limits: predicting capability gain does not establish autonomous
coarse dynamics or preservation of the matched null.

The adaptive-cellular pilot addresses a different component:
how local learning and inheritance affect population observables.
Changes in harvest and persistent behavioral types demonstrate
neither faster learning nor acquisition of new-task capability.
They motivate separating inherited solutions from inherited capacity
to learn, which remains untested here.

Together, these cases motivate preserving information placement,
task-dependent decisions, and the meaning of the null. They do not
supply additional complete RCI certificates or evidence of repeated
renormalization.

\section{Discussion}
RCI links an effective cell to the paired task experiment it preserves. The exact construction depends on symmetry and missing-role parity; it does not preserve arbitrary tasks or the joint distribution of all node answers. The frozen-model study covers one task family, one scale, and an adaptive candidate sequence. Positive relative-gain margin therefore supplies neither a complete reduction certificate nor an impossibility result.

Supporting probes expose descriptor and calibration limits. Six seed blocks give limited evidence across systems; nonnegative Gaussian propagation excludes harmful expected interaction. The inheritance pilot's finite type bins and single lattice size cannot establish criticality, open-ended evolution, or inherited learning capacity. Supplied rules and decoders also preclude a discovery claim.

No matched spectral, POD, or Koopman reduction baseline was run. The resource ledger measures state dimension, not deployment efficiency. The next theoretical requirement is composition with propagated errors, compatible nulls and interfaces, and preserved budget meaning; constructing such maps and separating inherited solutions from learning capacity remain hypotheses.

\section{Conclusion}
RCI specifies preservation of a task-conditioned interaction--null contrast across scales. Its analytical example establishes feasibility under explicit assumptions; the pooling counterexample establishes that closure alone is insufficient. Three frozen-model candidates instantiate the audit and retain all failures. The final candidate did not pass the full RCI certificate: its positive margin supports only $G_{\rm macro}>0.02$ under within-run uncertainty bounds, while closure, the stricter decision budget, and absolute-quality confidence requirements fail. Repeated renormalization, generality across cellular systems, and cross-generation learning capacity remain open.

\section*{Data availability and disclosure}
The September--October 2026 simulations are documented in \texttt{anc/rci\_supplement/}: historical summaries, frozen protocols, projection provenance, every reduction target-level success/failure table, five checkpoints, and verified training records. Evidence paths below are relative to that directory; manuscript images are in submission-root \texttt{figures/}. Original protocols and reproduction records remain archived (Appendix~\ref{app:provenance}). Analytical examples are symbolic derivations.

\paragraph{Limits of external recomputation.}
The packaged horizon CSV supports regenerating its summary curves. Verifying raw-array-to-summary calculations additionally requires the separate original v3 task inputs and five \texttt{evaluation\_arrays.npz} files, array ordering, normalization, and code. The source manifest records local paths and hashes; these identify files but are not download links or correctness proofs. Complete observation arrays, selected full states, and execution archives are not included. Full replay also needs original evaluation code, projection records, inputs, and runtime dependencies. No public full-array URL, repository, DOI, release, or independent external reproduction is claimed. Original-training sample-stream and software-version gaps are detailed in Appendix~\ref{app:training-origin}.

AI assistance supported exposition, analytical checks, LaTeX, evaluation implementation, monitoring, and auditing. The author-reported frozen-model evaluations involved no retraining. This revision reuses saved results without new model evaluation, training, or data generation; historical reproduction records refer to earlier executions. Responsibility remains with the authors.

\appendix
\section{Supporting inference model and descriptor calculations}\label{app:inference-details}
This appendix retains the detailed model-specific calculations underlying the supporting probes; they are separate from the learned localization reduction.

\subsection{A solvable complementary-information model}\label{sec:gaussian}
Let $A,B$ be independent standard normal latent variables and $Y=\sgn(A)\sgn(B)$. Each node belongs to a known role $r_i\in\{A,B\}$ and observes only its assigned latent plus independent unit Gaussian noise. Observations occupy two labeled channels; the unobserved channel starts at zero. Since the other latent sign is balanced and independent, $I(Y;o_i,r_i)=0$. The task requires combining complementary evidence rather than majority voting over independent estimates of $Y$.

Communication is $x(t+1)=Px(t)$ in each channel, with
\begin{equation}
 P=(1-\alpha)I+\alpha W,\qquad K=P^{T},\qquad 0\leq\alpha\leq1,\label{eq:propagation}
\end{equation}
where the exponent $T$ denotes the number of rounds, not transpose. On an undirected graph, $W_{ij}=1/\max(d_i,d_j)$ for an edge, with residual diagonal weight. Thus $P$ is nonnegative, symmetric, and stochastic. Each endpoint channel receives independent Gaussian readout noise of variance $\sigma^2>0$. The decoder multiplies its two channel signs. There is no terminal global voting stage. The self-copy null uses $K=I$ and has expected accuracy $1/2$.

For node $i$ and role $r$, define
\begin{equation}
 s_{ir}=\sum_{j:r_j=r}K_{ij},\qquad
 v_{ir}=\sum_{j:r_j=r}K_{ij}^{2}+\sigma^2,\qquad
 u_{ir}=s_{ir}/\sqrt{v_{ir}}.\label{eq:snr}
\end{equation}
Here $u_{ir}$ is an amplitude signal-to-noise ratio, distinct from the power ratio $u_{ir}^2$.

\begin{proposition}[Exact gain within the Gaussian model]\label{prop:gain}
Under independent latent variables, independent sensor noise, independent readout noise across channels, and the specified decoder,
\begin{equation}
 G_{\rm cap}=\frac{1}{2N}\sum_i r(u_{iA})r(u_{iB}),\qquad
 r(u)=\frac{2}{\pi}\arctan u.\label{eq:exact}
\end{equation}
In particular, $G_{\rm cap}\geq0$.
\end{proposition}
\begin{proof}
The endpoint in channel $A$ is $s_{iA}A+\zeta_{iA}$, where $\zeta_{iA}$ is independent Gaussian noise of variance $v_{iA}$. Its correlation with $A$ is $u_{iA}/\sqrt{1+u_{iA}^2}$. The Gaussian sign-correlation identity gives $\E[\sgn(A)\sgn(s_{iA}A+\zeta_{iA})]=(2/\pi)\arcsin(u_{iA}/\sqrt{1+u_{iA}^2})=r(u_{iA})$. Independence between channels makes the correlation with the parity target the product of their sign correlations. Binary accuracy is one half plus one half of that correlation. Averaging nodes and subtracting the null proves the result. Nonnegative weights imply $u_{ir}\geq0$.
\end{proof}

This is a model-specific calculation based on classical Gaussian identities, not a new general law of intelligence. It reveals a structural limitation: this family cannot test genuinely negative expected interaction gain. Readout noise prevents an arbitrarily weak, noiseless real-valued signal from carrying perfectly readable sign information.

\begin{proposition}[Finite-time locality constraint]\label{prop:local}
If node $i$ has no source of either one of the two roles within its $T$-step backward influence set, its expected accuracy is $1/2$. If this holds for every node, collective gain is zero. In the nonnegative Gaussian model, gain is positive whenever at least one node has positive total weights from both roles.
\end{proposition}
\begin{proof}
Local updates imply $K_{ij}=0$ outside the backward influence set. Absence of a role gives $s_{ir}=u_{ir}=0$, so its product in Eq.~\eqref{eq:exact} vanishes. If both role sums are positive, both sign correlations are strictly positive.
\end{proof}
This establishes an elementary reachability condition for the chosen task. It does not say that stronger coupling monotonically improves gain: stochastic averaging can change useful signal and noise in different ways. Nor does positivity define a phase transition.

\subsection{Candidate microscopic summaries and their failure modes}
Probe 1 used average off-diagonal response $\Lambda=N^{-1}\sum_i\sum_{j\neq i}|P_{ij}|$, a finite-horizon mixing descriptor $\mu_T=1-\mathrm{SLEM}(P)^T$, and an initial-observation complementarity quantity
\begin{equation}
 I_{\rm comp}=\frac{2n_A n_B}{N(N-1)}I(Y;U,V),
\end{equation}
where $U=A+\epsilon_A$ and $V=B+\epsilon_B$. This is a random two-node task-information quantity, not a full partial-information decomposition. $\Lambda$ is not a Lyapunov exponent. A fourth candidate used $n_{\rm eff}=N^{-1}\sum_i(\sum_jK_{ij}^2)^{-1}$.

None of these quantities identifies where the roles lie. Probe 2 instead formed
\begin{align}
 w_i&=\min(u_{iA},u_{iB}),&c_i&=\frac{w_i}{1+w_i},\\
 C&=\frac1N\sum_i c_i,&H&=\sqrt{\frac1N\sum_i(c_i-C)^2},\\
 B_{\rm imb}&=\frac1N\sum_i\frac{|u_{iA}-u_{iB}|}{1+u_{iA}+u_{iB}}.\label{eq:features}
\end{align}
The symbol $B_{\rm imb}$ denotes the archived feature named \texttt{B}, not the latent variable $B$. A ridge predictor uses $[C,C^2,C^3,CH,CB_{\rm imb},CHB_{\rm imb}]$ and clips predictions to $[0,1/2]$. The exact accuracy formula is excluded from its features. The descriptors nevertheless use substantial task knowledge: two labeled channels, their weaker-channel bottleneck, and noise calibration.

\begin{proposition}[Summary collision lower bound]\label{prop:collision}
If two systems or task distributions have the same descriptor $z$ but expected gains $g_1,g_2$, any deterministic predictor $f(z)$ has worst-case absolute error at least $|g_1-g_2|/2$ on that pair.
\end{proposition}
\begin{proof}
The triangle inequality gives $|g_1-g_2|\leq|g_1-f(z)|+|g_2-f(z)|$.
\end{proof}
The bound applies to the supplied descriptor, not to every possible microscopic representation. It distinguishes missing information from insufficient regression capacity.

\section{Additional control and literature comparisons}\label{app:position-tables}
These tables retain the original control hierarchy and conceptual positioning. They are not additional experiments or head-to-head empirical rankings.

\begin{table}[htbp]\centering\small
\caption{Control hierarchy. ``Observed'' refers to the archived studies; other rows are requirements for stronger claims, not completed experiments.}\label{tab:nulls}
\begin{tabularx}{\linewidth}{p{.22\linewidth}XX}
\toprule
Control & Preserved / changed & Evidence and interpretation\\
\midrule
Self-copy null & Same samples, states, read noise, decoder, deadline; $P=I$ & Observed in all inference probes; communication gain only.\\
Episode-misaligned channel & Endpoint channel marginal and spatial dependence; wrong task instance & Observed in Probe 1; checks task alignment, not a per-round fault process.\\
Role-placement pair & Graph, size, role counts, dynamics and noise; information locations & Observed; isolates location under fixed response mechanism.\\
Direct centralized agent & Full observation access and declared compute budget & Not measured; no superiority claim over centralized inference.\\
Serial same decomposition & Same subproblems, information, total work, and decoder & Not measured; no parallel-computation advantage claim.\\
Message/energy matched null & Equal traffic, precision, energy, and useful compute & Not measured; equal update weights are insufficient.\\
Random inheritance & Control-parent learning-vector norm; direction randomized & Observed in cellular pilot; trajectory-wide budgets can diverge.\\
No-learning inheritance & $\eta=0$, same seeds; vary inheritance strength & Observed in cellular pilot; implementation negative control.\\
\bottomrule
\end{tabularx}
\end{table}

\begin{table}[htbp]\centering\small
\caption{Concrete neighboring studies and the additional RCI preservation question. This is a conceptual comparison; no matched baseline experiment or claim of missing controls in those studies is implied.}\label{tab:recent-position}
\begin{tabularx}{\linewidth}{p{.22\linewidth}XX}\toprule
Study & Stated object or demonstration & Question posed by the present RCI contract\\\midrule
Dacre et al. (2025) & Local sensing and global geometric estimation in a physical sensor array & Does a reduced state preserve the declared task contrast under the same observation and null interfaces?\\
Schmidt et al. (2024) & Graph structure, spectra, and diffusion under coarse-graining/rescaling & Does preserved effective dynamics also retain decisions and the matched-null task score?\\
Etcheverry et al. (2026) & Cellular reasoning with local computation and asynchronous updates & Which state reductions preserve capability at a fixed task deadline and budget?\\
\bottomrule\end{tabularx}
\end{table}

\begin{table}[htbp]\centering\small
\caption{Relation of RCI to neighboring frameworks. The final column identifies the operational emphasis of this paper, without asserting that earlier theories cannot accommodate it.}\label{tab:position}
\begin{tabularx}{\linewidth}{p{.20\linewidth}XX}\toprule
Framework & Established object & RCI emphasis\\\midrule
Information bottleneck / information-theoretic RG & Relevant information and coarse degrees of freedom & Task decisions and an interaction--null contrast.\\
State abstraction / homomorphisms & Related dynamics, rewards, and value functions & Two experimental arms and their admissible observation interfaces.\\
Causal abstraction & Consistent predictions under mapped interventions & Communication removal with preserved decoder, weighting, and budget meaning.\\
Distributed inference & Propagation and collective estimation & Capability relative to local information access and a practical threshold.\\
\bottomrule\end{tabularx}
\end{table}
\section{Additional error metrics for continuous states}\label{app:continuous-errors}
\paragraph{When total variation is too strong.}
For deterministic continuous dynamics, different point-mass endpoints can have total variation one even when their Euclidean distance is small. Eq.~\eqref{eq:gainbound} is then valid but can be uninformative. One option is a task-conditioned Wasserstein comparison: if the two laws share the same target marginal and $\overline h_a(\cdot,y)$ is uniformly $L_a$-Lipschitz, then
\begin{equation}
 |G_{\rm micro}-G_{\rm macro}|
 \leq\eta_1+\eta_0+L_1\mathcal W_1+L_0\mathcal W_0,
 \label{eq:wasserstein}
\end{equation}
where $\mathcal W_a=\E_Y[W_1(\mu_a(\cdot\mid Y),\overline\mu_a(\cdot\mid Y))]$ and conditional laws are assumed to exist. This follows by coupling or Kantorovich--Rubinstein duality. It is not applicable to an arbitrary discontinuous decoder. Gaussian readout noise, however, smooths the parity decoder: for pre-readout channels $(z_A,z_B)$,
\begin{equation}
 \overline h_y(z_A,z_B)=\frac12\left[1+y\,
 \operatorname{erf}\!\left(\frac{z_A}{\sqrt2\sigma}\right)
 \operatorname{erf}\!\left(\frac{z_B}{\sqrt2\sigma}\right)\right].
 \label{eq:smoothedscore}
\end{equation}
Its Euclidean gradient norm is at most $1/(\sigma\sqrt\pi)$ for $\sigma>0$. This supplies a concrete setting for a Lipschitz comparison, while the bound deteriorates as readout noise vanishes. The historical inference cases do not estimate these metrics. Section~\ref{sec:endtoend} instead instantiates a direct coupled-output distance bound for a continuous localization loss; it does not estimate total variation or Wasserstein distance between full states.

\section{Detailed frozen reduction definitions and confidence calculations}\label{app:frozen-methods}
This appendix supplies the projection definitions, residuals, and confidence formula used in Section~\ref{sec:endtoend}.

\subsection{Three sequential candidates and their budgets}
The chronology and sample sizes are stated in Section~\ref{sec:endtoend}. Exact target-coordinate overlaps with accessible historical and preflight datasets were excluded; the complete historical training stream was not saved.

\paragraph{Block mean (v1).}
Let $L$ average each adjacent $2\times2$ block in all eight channels and $U$ repeat that state at the four original sites. The candidate update is $\overline F_\lambda=L F_\lambda U$, retaining the four raw observation ports. It stores 2,048 dynamic scalars rather than 8,192. Both the interaction and null roll autonomously. Although its two independently coded macro-null formulas agree numerically, repeated hidden-state averaging and broadcasting creates a within-block information path even at $\lambda=0$. Retaining separate raw observations alone therefore does not preserve the original null partition.

\paragraph{Four private dimensions per port (v2).}
Split the microscopic state into implemented output coordinates $y_i\in\R^2$ and hidden coordinates $h_i\in\R^6$. Let $W_h$ be the six hidden-state columns of $W_1$ and $V_2\in\R^{6\times2}$ its leading right singular vectors. Define $T_4=\operatorname{diag}(I_2,V_2^\top)$ and $E_4=T_4^\top$. The reduced state of a block contains the four separate vectors $T_4X_i$, rather than their average. Projection $P_4$ packs these labeled ports; lifting $U_4$ applies $E_4$ and restores their original positions. The update is $\overline F_{4,\lambda}=P_4F_\lambda U_4$. Its 4,096 dynamic scalars retain private histories, and its null is exactly port-factorized in the mathematical construction. Nevertheless, a low-rank approximation of $W_h$ is not a sufficient-state guarantee for nonlinear feedback.

\paragraph{One additional private residual mode (v3).}
The final candidate retains all four preceding coordinates and adds one hidden direction. Only pre-existing \emph{fit} trajectories supply its state statistics: 128 targets with four repeats, 64 deterministic sites, and times $0,4,8,16,24,32$, balanced across interaction and same-weight null. Define
\begin{equation}
 R_h=I-V_2(V_2^\top V_2)^{-1}V_2^\top,\quad
 C_\delta=\E_{\rm old\ fit}[(R_hh)(R_hh)^\top],\quad
 w=\operatorname{eigmax}(C_\delta),\quad
 T_5X=(y,V_2^\top h,w^\top h).
 \label{eq:residualmode}
\end{equation}
The direction is projected into the complement of $V_2$, normalized, and assigned a deterministic sign. The second moment is uncentered; no task labels or new formal-test outcomes select it. The same map serves both arms and all sites of a given model. The fifth coordinate is a genuine evolving private state, not a residual read from the microscopic trajectory during rollout. $E_5=T_5^\top$ and $\overline F_{5,\lambda}=P_5F_\lambda U_5$ give 5,120 dynamic scalars, a 37.5\% reduction, or a factor of 1.6 in dynamic-state capacity. No additional neural rule or decoder is fitted.

Table~\ref{tab:budget-ledger} records the unchanged observation ports, MLP work, and state counts; lifting and projection add computation.

\subsection{Closure, null consistency, and decision errors}
For arm $a\in\{1,0\}$, let $X_t^a$ be the autonomous microscopic path and $Z_t^a$ its autonomous reduced counterpart. Teacher states are used only to evaluate the residual
\begin{equation}
 r_a(t)=P F_a(X_t^a;O)-\overline F_a(PX_t^a;O).
 \label{eq:empclosure}
\end{equation}
They never replace $Z_t^a$. Per observation, $C_a=\max_{t<32}\max_j\|r_{a,j}(t)\|_2/(2\sqrt8)$; $j$ denotes a block for v1 and a labeled port for v2/v3. The denominator uses the original physical channel norm. We additionally retain raw maximum norms, per-coordinate RMS, and discarded-state energy. The closure estimands differ across these interfaces and projected dimensions; their numbers are not a paired same-estimand improvement factor.

Decision errors are directly coupled endpoint distances,
\begin{equation}
 D_a=\frac{\|q(X_{32}^a)-\overline q(Z_{32}^a)\|_2}{\sqrt2},\qquad
 |G_{\rm micro}-G_{\rm macro}|\leq\E D_1+\E D_0.
 \label{eq:directdecisionbound}
\end{equation}
The inequality follows from the reverse triangle inequality for the declared loss, separately in each arm. It instantiates a task-specific coupling bound without treating a small Euclidean error as a small total-variation error. $D_0$ remains necessary even when the reduced null's dependency graph is correctly factorized.

Implementation agreement, port isolation, and behavioral null approximation remain separate checks. In the private-port construction $PU=I$ and $q(X)=\overline q(PX)$ up to numerical tolerance; neither implies $Z_t=PX_t$.

\subsection{Frozen inferential contract}
For each target we first average the four repeats and the five fixed models. The 1,024 target-level values are the independent units under the specified sampling model. For a bounded statistic of range width $R$, unbiased sample variance $s_n^2$, and $\beta=0.05/6$, we use the scaled empirical Bernstein radius \citep{maurer2009}
\begin{equation}
 r_R=\sqrt{\frac{2s_n^2\log(2/\beta)}{n}}+
 \frac{7R\log(2/\beta)}{3(n-1)},\qquad n=1024.
 \label{eq:bernstein}
\end{equation}
$G_{\rm micro}$ has width two; $D_1,D_0,C_1,C_0$ and the macro-interaction loss have width one. We use a lower bound $G_L$ for gain and upper bounds for the remaining five statistics, with upper clipping at one where appropriate. A union bound gives a within-run simultaneous 95\% coverage statement for these six directions, conditional on the already fixed collection and projection. This is not a simultaneous 95\% guarantee over all three adaptively developed candidates. Supplementary per-model bounds allocate $0.05/30$ and never select the reported model collection.

\subsection{What these tests establish about RCI}
A correct projection readout is weaker than preservation of future decisions. For private-port channel projections $\Pi=EP$, suppressing packing notation, $\Pi$ commutes with the linear spatial operator $A$. The teacher residual still contains
\begin{equation}
 r_\lambda(X)=\tfrac12P\{\phi_\theta(B_\lambda X,O)-
 \phi_\theta(B_\lambda\Pi X,O)\},\qquad B_\lambda=(1-\lambda)I+\lambda A.
 \label{eq:channeldefect}
\end{equation}
The removed hidden components change nonlinear updates despite unchanged current output coordinates. The coupled state error obeys
\begin{equation}
 PX_{t+1}-Z_{t+1}=r_t+\{\overline F(PX_t)-\overline F(Z_t)\}.
\end{equation}
This separates omitted-state closure from subsequent feedback. In v2 the first-step projection is exact, yet discrepancy starts with the next update and the final interaction decision distance averages 0.1755. Its point-estimate margin is already negative before confidence penalties, so more test samples alone would not repair that candidate.

On old balanced fit states, the added direction explains 54--83\% of omitted energy; on unchanged select/check states it removes 53--88\% across models and arms. These are geometry diagnostics, not rollout or causal effects. The gates in Eq.~\eqref{eq:allgates} determine preservation.

\section{Archived supporting empirical case studies}\label{app:case-details}
\subsection{Role and statistical interpretation}
Table~\ref{tab:caseroles} distinguishes these supporting studies from the separate state-reduction experiment.

\begin{table}[htbp]\centering\small
\caption{What the supporting cases contribute. None of the four historical supporting cases in this table tested a dynamical state coarse-graining map; the separate frozen reduction tests appear in Section~\ref{sec:endtoend}.}\label{tab:caseroles}
\begin{tabularx}{\linewidth}{p{.18\linewidth}XX}\toprule
Case & Supported mechanism or diagnostic & Relation to RCI\\\midrule
Inference 1 & Information placement and task alignment matter beyond general propagation summaries & Identifies distinctions a useful description must retain.\\
Inference 2 & Co-reachability predicts within-family gain under frozen evaluation & Supports task-conditioned response compression.\\
Inference 3 & Finite calibration works; response-moment transfer fails across distributions & Separates measurability from descriptor sufficiency.\\
Adaptive cellular pilot & Inheritance changes harvest and persistent behavioral novelty & Motivates adaptive effective units; does not measure new-task capability.\\
\bottomrule\end{tabularx}
\end{table}

The sequential probes use separate discovery and frozen testing within each study. Counts are 3,120 condition--seed runs in Probe 1, 4,368 in Probe 2, 1,872 task runs in Probe 3, and 180 cellular trajectories. Probes 1/2 use 1,024 episodes per run; Probe 3 uses 2,048. Its four calibration predictions yield 7,488 rows, not independent task datasets. Shared random numbers require seed-block uncertainty summaries; intervals are descriptive, without pooling these studies or confirmatory multiplicity claims.

\subsection{Probe 1: useful gain is not agreement or topology alone}
The first probe used ring, periodic grid, small-world, modular, and star graphs with $N\in\{16,32,64,128\}$; $\alpha\in\{0,.1,.3,.6,1\}$; and $B$-role proportions $p\in\{0,.125,.25,.5\}$. Communication lasted eight rounds with $\sigma=0.2$. Six seeds were used per configuration. Ring/grid systems of size 16 and 32 supplied 480 discovery runs; 1,920 subsequent runs tested new sizes, new topologies, or both. Another 720 runs tested failure controls. Regression coefficients and code hashes were frozen before held-out generation. Graph degree, message count, and energy were not all matched; in particular, the star is not a bounded-degree network family.

The three-variable predictor had condition-mean RMSEs of 0.03324, 0.02405, and 0.02692 on size, topology, and joint transfer. The size error missed the prespecified 0.03 criterion. A candidate augmented with weight concentration attained errors 0.00867, 0.01107, and 0.01899, respectively. Both results include easy zero-communication and zero-complementarity conditions. A post-hoc active-condition diagnostic reduced the three-variable size-transfer $R^2$ to 0.359; this diagnostic is not a replacement primary analysis.

More decisively, a modular graph with $N=128$, $\alpha=1$, and $p=.5$ produced the results in Table~\ref{tab:collision}. Random versus segregated roles preserve the three scalar descriptors and weight concentration, yet change whether complementary information can meet by the deadline. Their paired gain difference was 0.2118, with a six-seed block-bootstrap descriptive interval $[0.2071,0.2162]$. The corresponding analytical difference was approximately 0.2105. Proposition~\ref{prop:collision} therefore gives a worst-case descriptor-only prediction error of at least approximately 0.1053 for that pair.

\begin{table}[htbp]\centering
\caption{Archived Probe 1 stress test. Intervals are descriptive 95\% seed-block bootstrap intervals, not multiplicity-adjusted tests.}\label{tab:collision}
\begin{tabular}{lcc}\toprule
Condition & Capability gain & Agreement\\\midrule
Random roles & $0.2284\ [0.2234,0.2332]$ & 0.6590\\
Segregated roles & $0.0166\ [0.0161,0.0170]$ & 0.5011\\
Random roles, misaligned channel & $0.0016\ [-0.0022,0.0065]$ & 0.6596\\
\bottomrule\end{tabular}
\end{table}

Cyclically shifting the terminal $B$ channel across episodes preserves its empirical marginal and spatial dependence while breaking target alignment. Its near-original agreement and near-zero gain separate agreement from useful integration. This endpoint intervention is not an equal-cost alternative communication protocol.

\begin{figure}[htbp]\centering
\IfFileExists{figures/rci_probe.png}{\includegraphics[width=\linewidth]{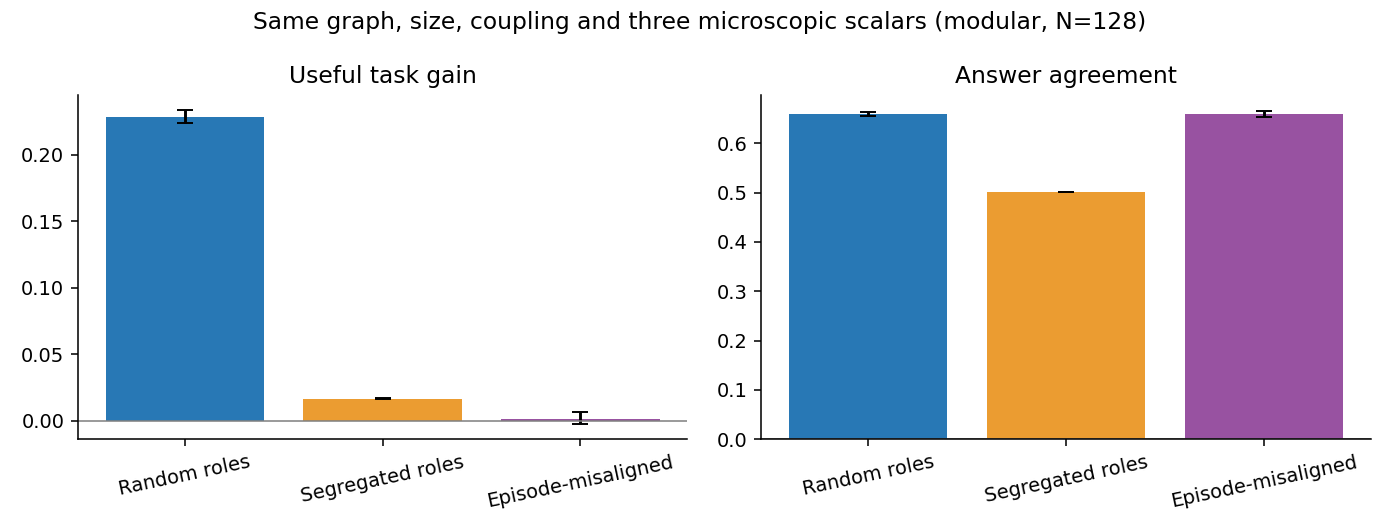}}{\fbox{\parbox{.88\linewidth}{\centering Figure supplied with the source package: \texttt{figures/rci\_probe.png}. Numerical results are included in the tables and text.}}}
\caption{Probe 1: information alignment and task gain. Left: task gain. Right: agreement. Error bars are the descriptive 95\% seed-block bootstrap intervals. Spatial role segregation and episode misalignment have different effects on agreement despite both reducing useful gain. The plot's reference to ``three microscopic scalars'' concerns the original descriptors, not all possible task-aware information measures.}\label{fig:probe1}
\end{figure}

\subsection{Probe 2: task-aligned co-reachability predicts within-family transfer}
The second probe fixed the descriptors in Eq.~\eqref{eq:features} before testing. Discovery used 192 configurations with six seeds, giving 1,152 runs. Configurations varied ring/grid and one- or four-bridge modular graphs, $N=32/64$, random/segregated roles, $T=4/8$, and readout noise $0.1/0.3$. Coupling was $0.15/0.45/0.85$ and role proportion $0.25/0.5$. Regularization was selected only within discovery, leaving out topology--layout groups. Discovery and test seed sets were disjoint.

Five held-out groups each contained 96 configurations and six seeds: new block/smooth layouts; $T=2/16$; noise $0.05/0.6$; new matching/eight-bridge graphs at $N=128$; and joint changes. Maximum graph degree was five. The remaining 336 test runs comprised 288 layout-repair runs and 48 null checks. The repair subset contains 48 configurations, forming 24 random--segregated pairs; its count erratum is documented in Appendix~\ref{app:provenance}.

\begin{table}[htbp]\centering
\caption{Probe 2 held-out metrics from \texttt{results/metrics.csv}; configuration means over six seeds. The older descriptor structure was refitted on this probe's discovery data, not reused with Probe 1 coefficients.}\label{tab:probe2}
\begin{tabular}{lrrr}\toprule
Held-out change & Old-descriptor RMSE & Co-reachability RMSE & $R^2$\\\midrule
Layout & 0.05758 & 0.00355 & 0.9975\\
Horizon & 0.06378 & 0.00345 & 0.9984\\
Noise & 0.09599 & 0.00557 & 0.9967\\
Topology and size & 0.07378 & 0.00283 & 0.9989\\
Joint & 0.10843 & 0.00951 & 0.9932\\
\bottomrule\end{tabular}
\end{table}

Pooled RMSE fell from 0.08222 for the older descriptor structure to 0.00555 for co-reachability, a 93.2\% reduction. A one-variable $C$ predictor attained 0.01829. The three-statistic predictor passed all five archived RMSE/$R^2$ gates, the pooled improvement gate, and the repair-pair gate. Random-minus-segregated gain differences had prediction RMSE 0.00697, compared with 0.14115 for the old descriptors. At the operational $G>0.02$ threshold, pooled precision was 1.000 and recall 0.9956.

These are average-error results, not uniform guarantees. The largest joint-group absolute error was 0.04030, exceeding 0.03. Furthermore, the new descriptors add noise calibration as well as location information; the 93.2\% reduction is not an isolated causal estimate of a spatial contribution. The paired-layout comparison is the more direct evidence about location. Probe 2 does not repair the episode-misalignment counterexample because it assumes correctly aligned, reliable channels.

\begin{figure}[htbp]\centering
\IfFileExists{figures/rci_probe2.png}{\includegraphics[width=\linewidth]{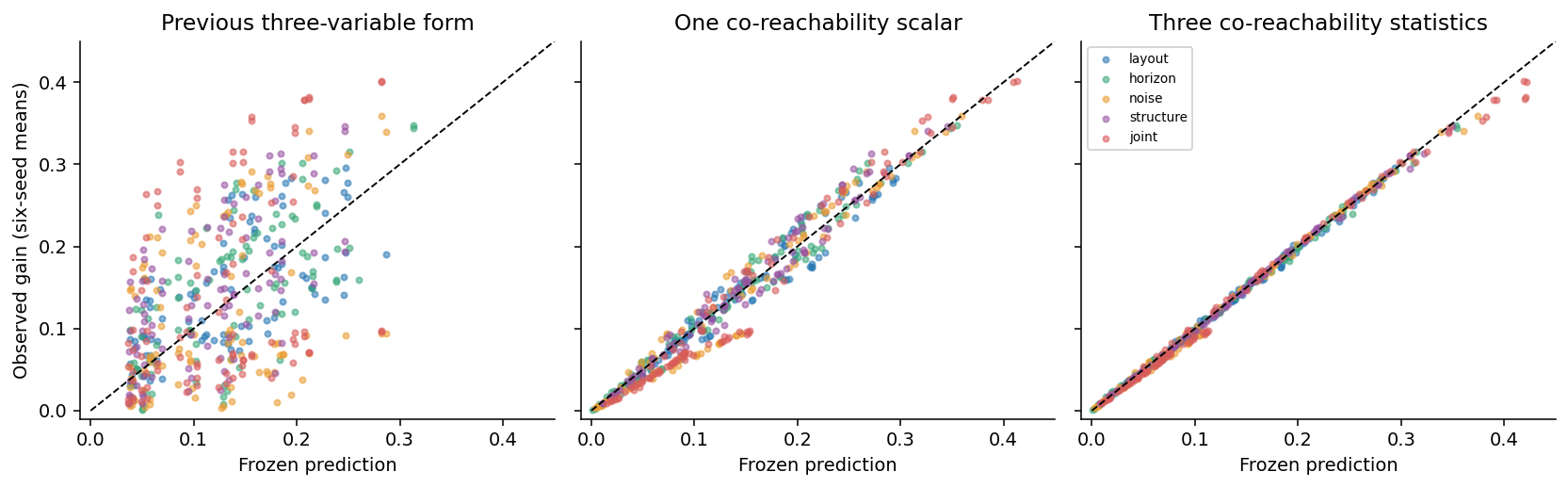}}{\fbox{\parbox{.88\linewidth}{\centering Figure supplied with the source package: \texttt{figures/rci\_probe2.png}. Numerical results are included in the tables and text.}}}
\caption{Probe 2: prediction from co-reachability. Frozen predictions versus observed configuration-mean gains across five held-out groups. Three co-reachability statistics improve on the older descriptor structure and a single co-reachability scalar within the linear Gaussian family. Points are configurations averaged over six seeds, not independent experiments for uncertainty estimation.}\label{fig:probe2}
\end{figure}

\subsection{Probe 3: finite calibration succeeds; distribution transfer fails}
Probe 3 retained Probe 2's six coefficients without refitting. It used 96 main configurations, eight control configurations, six new seeds, and three environments. Calibration applied known positive and negative unit stimuli to both channels and recorded noisy endpoints. The estimator used stimulus signs and endpoint responses, not the true matrix $K$, noise parameters, node-role map, or task labels. Signal was estimated by half the positive-minus-negative mean difference, truncated below at zero; noise was estimated by pooled within-stimulus variance. Calibration budgets 32, 128, and 512 were nested prefixes of the same independent calibration stream. Although rounds do not scale with $N$, total observations scale as $O(RN)$, and controllable stimuli plus access to every endpoint remain strong instrumentation assumptions.

The three environments retained the same parity target and decoder: (i) Gaussian latents plus Gaussian sensor noise; (ii) equiprobable $\pm1$ latents plus Gaussian noise; (iii) $\pm1$ latents with binary observations $\pm\sqrt2$ and flip probability $(1-1/\sqrt2)/2$. Under unit stimuli, their conditional response means and variances match. Consequently, ideal $(C,H,B_{\rm imb})$ coincide configuration by configuration. The environments need not have equal task difficulty; this is precisely a test of information missing from response moments.

\begin{table}[htbp]\centering
\caption{Probe 3 RMSE from the result table, evaluated on 96 main configuration means over six seeds. Ideal means true response parameters, not zero-data estimation. All columns use the same frozen predictor.}\label{tab:probe3}
\begin{tabular}{lrrr}\toprule
Calibration & Gaussian & $\pm1$ latent + Gaussian noise & Binary observation\\\midrule
Ideal & 0.00358 & 0.13114 & 0.12215\\
32 rounds & 0.00575 & 0.13164 & 0.12011\\
128 rounds & 0.00359 & 0.13104 & 0.12252\\
512 rounds & 0.00345 & 0.13129 & 0.12118\\
\bottomrule\end{tabular}
\end{table}

At the primary 512-round budget, Gaussian $R^2$ was 0.9981 and the descriptive seed-block RMSE interval was $[0.00337,0.00443]$. All prespecified calibration gates passed. Run-level RMSE was 0.00487, rather than the lower configuration-mean error, and 32-round run-level RMSE was 0.01012. Averaging seeds must therefore not be presented as the error of a single deployed calibration. A slightly lower mean error than ideal calibration reflects sampling and error cancellation, not a benefit of adding noise.

Both ideal and 512-round binary transfer failed the RMSE/$R^2$ gates. At 512 rounds, binary $R^2$ was only 0.0645. Increasing calibration budget did not remove the distribution error. For a representative ring configuration ($N=32$, random roles, $T=16$, $\sigma=.1$, $\alpha=.3$, $p=.5$), empirical gains were 0.26783, 0.47046, and 0.44280 in the three environments, while the shared ideal frozen prediction was 0.26382. The analytical expected gap between the first two environments was 0.20551. The descriptor collision thus forces at least approximately 0.10276 absolute error in one of them for any deterministic descriptor-only predictor.

The cause can be derived. With $\pm1$ latents and Gaussian endpoint noise, channel sign correlation is $r_{\pm}(u)=2\Phi(u)-1$; with Gaussian latents it is $r_G(u)=2\arctan(u)/\pi$. Both depend on the same unit-stimulus ratio $u$, but differ because probability mass near the decision boundary differs. For independent channels, the gain remains one half of the average product of the appropriate correlations. Thus a useful factorization separates propagation response from the task-distribution-dependent decision map. The third environment also changes the noise law, for which second moments alone need not determine sign error.

\begin{figure}[htbp]\centering
\IfFileExists{figures/rci_probe3.png}{\includegraphics[width=.88\linewidth]{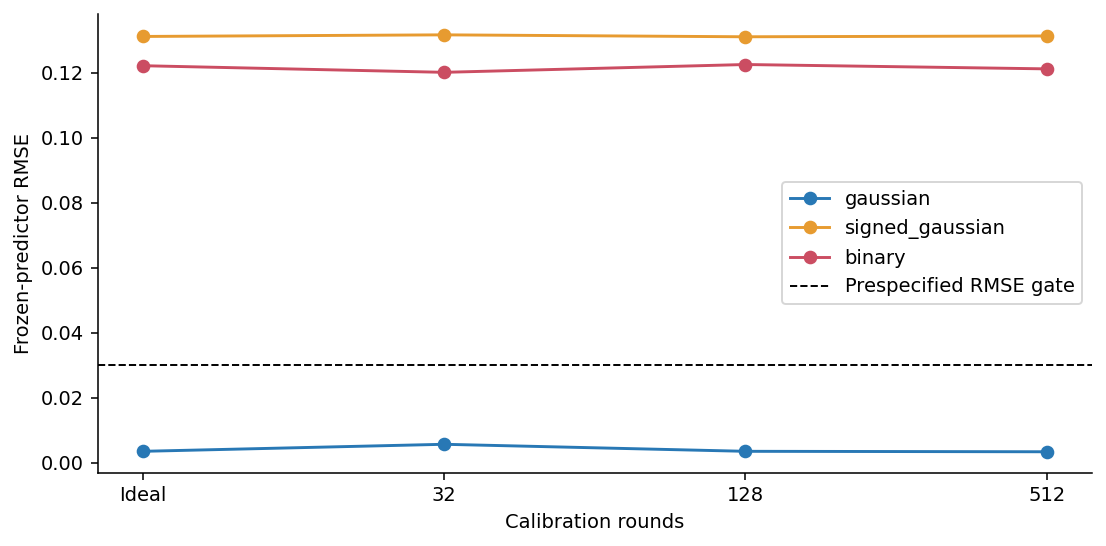}}{\fbox{\parbox{.88\linewidth}{\centering Figure supplied with the source package: \texttt{figures/rci\_probe3.png}. Numerical results are included in the tables and text.}}}
\caption{Probe 3: calibration and distribution transfer. Calibration improves or preserves accuracy within the Gaussian family but does not repair distribution transfer. The dashed line is the prespecified RMSE gate of 0.03, not a statistical significance threshold.}\label{fig:probe3}
\end{figure}

\subsection{Adaptive cellular pilot: inheritance changes proxies, not demonstrated capability}
The separate cellular pilot ran a $64\times64$ periodic Moore-neighborhood lattice for 10,000 sweeps. Each sweep restores resources, then makes $64^2$ asynchronous site-update attempts with replacement. Cells use a three-action linear-softmax policy (rest, harvest, divide) with six local features and 18 parameters. Genotype $g$ initializes the learned policy $\theta$ at birth. A local harvest-reward gradient updates $\theta$; inheritance is
\begin{equation}
 g_{\rm child}=g_{\rm parent}+\lambda_A(\theta_{\rm parent}-g_{\rm parent})+\epsilon,
 \qquad \epsilon\sim\mathcal N(0,\mu^2I).\label{eq:inherit}
\end{equation}
The mutation parameter is a per-coordinate standard deviation, not mutation probability. Every birth mutates. There is no global fitness reward. Local costs, background mortality, and resource replenishment make population persistence endogenous.

The design comprised 75 main runs ($\eta=.05$, five $\lambda_A$ values, three mutation levels, five seeds), 30 randomized-direction writeback controls at $\lambda_A=.5/1$, and 75 no-learning controls. All runs used the same horizon, with no outcome-based early stopping. A presimulation feasibility check changed resource replenishment from .06 to .12 after a low-resource population decline; this disclosed calibration means the pilot is not a wholly untouched preregistration.

Early performance $A_{\rm short}$ is the mean of ten harvest-per-action records at sweeps 20--200. Persistent novelty counts newly observed genotype-defined behavior types that meet abundance, duration, and generational-depth conditions (Appendix~\ref{app:cell}). The late novelty slope $v_N$ is measured over sweeps 5,000--10,000 in types per 1,000 sweeps. A lagged mutual-information proxy measures predictive persistence, not established organizational complexity.

\begin{table}[htbp]\centering
\caption{Selected cellular pilot observations. Intervals are descriptive 95\% $t$ intervals over five seeds. The maxima are selected within the tested grid, not population-optimal parameter estimates.}\label{tab:ca}
\begin{tabularx}{\linewidth}{Xcc}\toprule
Quantity & $(\lambda_A,\mu)$ & Mean [interval]\\\midrule
Maximum early harvest & $(1,.01)$ & $0.122861\ [0.122683,0.123038]$\\
Maximum late novelty rate & $(0,.08)$ & $6.415\ [5.130,7.699]$\\
Learned minus random early harvest & $(1,.01)$ & $0.002474\ [0.002430,0.002517]$\\
Learned minus random late novelty & $(1,.01)$ & $-13.99\ [-16.15,-11.83]$\\
\bottomrule\end{tabularx}
\end{table}

The best early-harvest point exceeded the $\lambda_A=0,\mu=.01$ mean by approximately 2.01\%. This is a modest finite-time performance difference, not demonstrated acceleration of learning. In fact, all main-condition early harvest curves declined from the high-resource initial state, and the $T_{90}$ measure took its earliest available value of 60 sweeps throughout. It cannot rank adaptation speeds here.

At $\lambda_A\geq.5$, no seed at the two lower mutation levels added a new persistent type in the second half, whereas all seeds at $\mu=.4$ did. This is an exploratory finite-horizon novelty/saturation classification. Main-condition late occupancy remained approximately .882--.999, crossing none of the prespecified population thresholds .2/.5/.8; all 180 trajectories ended nonextinct. There was no observed population-boundary crossing, multi-size critical scaling, or evidence of open-ended evolution.

The randomized direction preserves the norm of the \emph{control trajectory's own} parental learning displacement at each birth. Once trajectories diverge, their total displacement distributions and birth histories can differ. The control therefore does not isolate semantic content under globally equal perturbation budgets. Its larger novelty rate in some settings further demonstrates that additional types can reflect undirected exploration rather than useful adaptation. With $\eta=0$, all five $\lambda_A$ trajectories and terminal states were identical within each mutation--seed pair in the saved validation: the learning-writeback channel vanishes as intended.

\begin{figure}[htbp]\centering
\IfFileExists{figures/ca_controls.png}{\includegraphics[width=\linewidth]{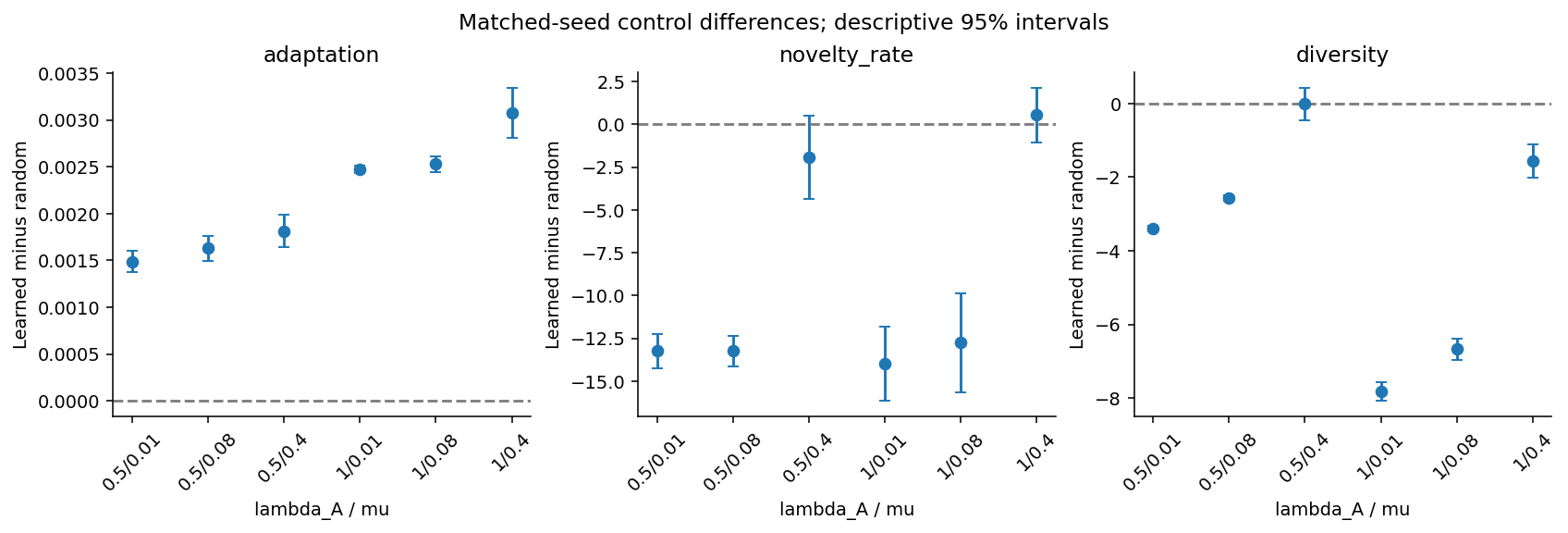}}{\fbox{\parbox{.88\linewidth}{\centering Figure supplied with the source package: \texttt{figures/ca\_controls.png}. Numerical results are included in the tables and text.}}}
\caption{Adaptive cellular pilot: direction-randomized inheritance control. The axis label ``adaptation'' denotes early mean harvest, not learning speed; ``novelty\_rate'' denotes persistent-type slope, not new-task ability; ``diversity'' denotes genotype distance, including potentially neutral directions. Error bars are descriptive 95\% intervals for five paired seeds. Norm matching is within each control-parent event, not across complete treatment trajectories.}\label{fig:ca}
\end{figure}

The cellular study does not include a matched non-interaction task score, an independent new-environment challenge, or a state-reduction test. Its observations therefore concern adaptive population dynamics and behavioral descriptors, rather than a measured capability-onset boundary under Eq.~\eqref{eq:gain}.

\section{Reproduction details and evidence map}\label{app:provenance}
\subsection{Probe predictors and calibration}
Probe 1 selected ridge regularization from $\{10^{-6},10^{-4},.01,1\}$ using seed-grouped discovery cross-validation. Its three-variable feature basis was $[I\Lambda,I\Lambda^2,I\mu_T,I\mu_T^2,I\Lambda\mu_T]$. The concentration candidate used $[I\Lambda,I\Lambda^2,I\mu_T,I\Lambda\mu_T,I\log n_{\rm eff},I(\log n_{\rm eff})^2]$. The two-sensor mutual information entering $I$ was evaluated by 48-order Gauss--Hermite integration, independently of downstream accuracy.

Probe 2 used the same regularization grid with topology--layout grouped validation. Its old-descriptor comparator omitted the constant two-sensor information factor, using $2n_A n_B/[N(N-1)]$; this factor is common within the fixed observation family and can be absorbed by coefficients. The main three-statistic feature basis appears in the main text. Alternative fixed candidates were $[C,C^2,C^3]$ and three coverage fractions $\Pr(w>.25)$, $\Pr(w>.75)$, $\Pr(w>1.5)$. All main evaluation groups excluded $\alpha=0$ and $p=0$, with those checks retained separately.

For Probe 3, if $z_{irk}$ is endpoint response at node $i$, channel $r$, and calibration stimulus sign $a_k$, the estimator computes
\begin{equation}
 \widehat s_{ir}=\max\{(\overline z_{ir,+}-\overline z_{ir,-})/2,0\},
 \qquad \widehat u_{ir}=\frac{\widehat s_{ir}}{\sqrt{\max(\widehat v_{ir},10^{-8})}},
\end{equation}
where $\widehat v$ pools the two within-sign sample variances with degrees-of-freedom weights. The positive truncation can give a positive estimated response for a true no-signal channel; no post-hoc correction was introduced. The main calibration gate used 512 rounds, while 32 and 128 were prespecified secondary budgets. The binary-transfer gate required both ideal and 512-round RMSE at most .03 and $R^2$ at least .8; both failed.

\subsection{Cellular implementation and measurements}\label{app:cell}
Initial site occupancy is independent with probability .25, initial energy is 1.5, initial genotype coordinates are $\mathcal N(0,.1^2)$, and resources start at 1. The six features are
\begin{equation}
 f=(1,E/4,R,\rho_{\mathcal N},\overline R_{\mathcal N},h_{\rm last}/.5),
\end{equation}
where $\rho_{\mathcal N}$ is neighboring occupancy, $\overline R_{\mathcal N}$ is neighboring mean resource, and $h_{\rm last}$ is the preceding harvest. Resource capacity is 1 and replenishment is .12 per sweep. A harvest takes at most .5 resource; action costs for rest/harvest/divide are .04/.08/.12. Energy is capped at 4; nonpositive energy causes death. After a divide action, energy above 2 permits reproduction into a random empty neighboring site, splitting remaining energy equally. Background mortality is .002 per selected-cell update. Newborns may be selected later in the same sweep. The learning update is $\theta\leftarrow\theta+\eta r_{\rm harvest}\nabla_\theta\log\pi_\theta(a\mid o)$, with birth initialization $\theta=g$.

Genotype-defined behavioral types use four fixed probe feature vectors $[1,E,R,.5,.5,0]$, $E\in\{.3,.8\}$ and $R\in\{.2,.8\}$. Harvest and divide probabilities are each discretized into three bins, yielding eight ternary digits and at most 6,561 types. A new type must have at least three individuals at four consecutive 20-sweep samples, spanning 60 sweeps, while its maximum generational depth increases by at least two. Initially supported types are excluded. This criterion does not trace a single ancestral lineage. A two-bin sensitivity analysis is also archived. Finite type capacity alone permits eventual saturation.

Predictive persistence is
\begin{equation}
 C_{\rm pers}=\max\left\{0,I(S_i(t);S_i(t+20))-
 \frac18\sum_{k=1}^8 I(S_i(t);S_{i+\Delta_k}(t+20))\right\},
\end{equation}
using site states empty/rest/harvest/divide and nonzero spatial-shift baselines. It is distinct from the inference feature $C$ and can reflect density and longevity. Genotype diversity is the mean Euclidean distance of 512 sampled distinct-individual pairs; softmax-invariant directions can contribute neutral variation.

\subsection{Protocol chronology and historical verification}
Probe 2 used discovery seeds 101--106 and test seeds 201--206. Probe 3 used seeds 301--306. The original Probe 2 protocol contained a repair-subset count typo: the frozen design has 48 configurations, not 96. The original protocol and a separate erratum were retained, with no deletion of experimental configurations.

Historical reruns reproduced Probes 1/2 CSVs, paired arrays, and coefficients, and Probe 3's six CSVs and task arrays. Cellular checks cover completeness, finiteness, resource/energy bounds, novelty monotonicity, no-learning invariance, randomized-direction norm preservation, a gradient finite-difference check, and one exact replay. These earlier implementation checks were not rerun here and do not validate scientific interpretations.

Earlier manuscript audits checked source tables and trajectories, including 180 early-harvest summaries. The provenance indexes connect historical evidence, reductions, and training; original protocols and errata remain archived. The four historical figures are unchanged exports, and Figure~\ref{fig:v3-errors} relayouts frozen statistics. Captions qualify broader source labels.

\begin{table}[htbp]\centering\small
\caption{Claim-to-artifact map within the ancillary evidence. Paths are relative to \texttt{anc/rci\_supplement/evidence/}.}\label{tab:evidence}
\begin{tabularx}{\linewidth}{p{.23\linewidth}Xp{.25\linewidth}}\toprule
Claim & Main artifact & Status\\\midrule
Topology-summary collision & \texttt{rci\_probe/results/summary.json}; \texttt{stress.csv} & Observed and analytically diagnosed\\
Frozen co-reachability transfer & \texttt{rci\_probe2/results/metrics.csv}; \texttt{frozen.json} & Observed within Gaussian family\\
Finite calibration / failed transfer & \texttt{rci\_probe3/results/metrics.csv}; \texttt{matched\_moment\_pairs.csv} & Observed, no refit\\
Inheritance and novelty & \texttt{paper1\_pilot/condition\_summary.csv}; \texttt{paired\_contrasts.csv} & Exploratory finite-time proxies\\
Analytical reduction & Section~\ref{sec:analytic} and its derivation & Exact symbolic result\\
Frozen cellular state reductions & \texttt{reductions/} (one directory per candidate), \texttt{primary\_result.json} & Evaluated; all candidates retained\\
Original model training & \texttt{training/verified\_training\_metadata.json}; checkpoint files and source hashes & Archived records verified; no retraining\\
Multilevel RG composition & No measured artifact & Not evaluated\\
Centralized/serial superiority & No measured artifact & Not evaluated\\
Cellular new-task capability & No measured artifact & Not evaluated\\
\bottomrule\end{tabularx}
\end{table}

\section{Open questions and falsification criteria}
The following questions delimit the reach of the framework beyond the analytical example, supporting case studies, and single-level learned-model evaluations. They are prospective research questions rather than additional empirical results.
\begin{enumerate}
\item \textbf{Coarse dynamics beyond the analytical construction:} specify $\pi_b$, $\tau_b$, the macrostate, the fitted or derived kernel, and the held-out trajectory split. Report closure error and decoder error separately for both arms. Failure of a memoryless closure may motivate a memory-bearing macrostate, but this is a new model requiring a new test.
\item \textbf{Capability-onset uncertainty:} freeze $\delta$, the tested condition family, and a simultaneous uncertainty procedure. Report unresolved boundary intervals rather than interpolating a critical point from a sparse grid.
\item \textbf{Distribution-aware transfer:} predefine what information about the task distribution is supplied, its measurement cost, and which mapping is frozen. Report zero-shot transfer separately from refitting. The present moment-matched failures remain part of the record.
\item \textbf{Adaptive cellular capability:} specify an external challenge and score before selecting a favorable novelty regime. Include appropriate no-learning and non-interaction controls, and distinguish general environmental robustness from a learned new task.
\end{enumerate}

\section{Frozen reduction provenance and additional diagnostics}\label{app:reduction-provenance}
The three formal streams use target seeds 2026102402, 2026102502, and 2026102602; preflight seeds are 2026102401, 2026102501, and 2026102601. All contain 1,024 formal targets and four repeats. Preflight uses 32 targets only for implementation and timing checks. Original weights and all formal code hashes are recorded in each frozen manifest. Matrix derivation for the fifth dimension records the historical fit-source hashes and residual eigenspectra. The fifth-coordinate basis is not selected from formal v1/v2/v3 scores. The \emph{decision to test that candidate} was informed by preceding failures and is not described as untouched preregistration.

All main statistics are recomputed from saved observation arrays and checked against per-target tables at tolerance $2\times10^{-7}$. Checkpoints are bytewise checked before and after execution. Every run retains 5,120 target--model rows and full output sequences; private-port runs additionally export all 20,480 observation--model rows. Descriptive case failure flags use the frozen numerical tolerances without per-case inferential claims. Selected complete-state snapshots cover the first four targets and all their repeats at steps $0,8,16,24,32$; this small subset is not treated as representative sampling of the task distribution.

For v2, the final microscopic loss is 0.098094 under interaction and 0.354267 under the same-weight null; the reduced losses are 0.219509 and 0.385268. Mean gain remains positive but the unpenalized point-estimate margin $\widehat G-0.02-\widehat D_1-\widehat D_0=-0.041235$ is already negative. More formal targets alone would not remove its actual endpoint discrepancy. Splitting each model's squared decision error into mean bias plus residual variation attributes 77.0\% of pooled interaction squared error and 99.66\% of pooled null squared error to model-specific mean offsets. These are descriptive identities, not interventions proving an attractor mechanism or justification for fitting a new terminal decoder on exposed test data.

The historical residual-mode analysis uses only saved states and fixed-matrix operations. A single constant omitted-state mean explains only 1.7--38.5\% of the balanced fit residual energy, whereas its leading uncentered mode explains 54--83\%. Reconstruction improvement after adding a coordinate is not a theorem that projected transition errors or autonomous decision errors decrease. A larger projection also measures an extra component in the closure residual. All claims about these quantities therefore use the separately frozen rollout results in Section~\ref{sec:endtoend}.

\section{Verified origin of the five frozen learned models}\label{app:training-origin}
The models come from an earlier communication-dependence study whose archived cohort contains seeds 1000--1019. The present RCI evaluations reuse the fixed subset 1000--1004; they do not pool the other models into the analysis or claim that these five are a random sample of all possible training seeds. This subsection reports historical records, not newly run training or validation.

\paragraph{Objective and update budget.}
The archived configuration specifies the same $32\times32$ lattice, eight state channels, 32-unit hidden layer, 32 inference updates, residual factor 0.5, communication strength 0.5, observation probability 0.2, noise standard deviation 0.1, Gaussian width 0.1875, and coordinate inputs used in the microscopic evaluation. A global mean of the first two normalized channels predicts the target. Supervised training minimizes the mean squared coordinate error over the batch and both coordinates by backpropagation through the entire 32-step unroll. The fingerprint-matched training loop uses Adam with learning rate $10^{-3}$, $\beta_1=0.9$, $\beta_2=0.999$, $\epsilon=10^{-8}$, zero weight decay, and gradient-norm clipping at 1.0. No learning-rate scheduler is present in that loop. Every archived run completes 24,000 optimizer updates with batch size 8, or 192,000 example draws per seed; these are draws, not a claim of that many unique archived samples. The configuration records CPU execution with one thread. Locality is a property of the inference rule, not a claim that the original training signal or weight updates were decentralized.

\paragraph{Random streams and checkpoint selection.}
Initialization seeds are 1000--1004. The source seeds Python, NumPy, and Torch from the model seed, requests deterministic Torch algorithms, and seeds the training-data generator with $1000000007+\text{seed}$. The common historical validation set has 256 targets with generator seed 170000003. Mean normalized Euclidean validation error is checked every 100 updates; the saved \texttt{best\_model} is the strict minimum over updates $100,200,\ldots,24000$. Step zero is logged but is not assigned as the selected best checkpoint. Training continues to the fixed budget rather than stopping at the selected step. Table~\ref{tab:original-checkpoints} gives the selected steps and historical validation errors. The older study also records a 4,096-target test stream with seed 190000003, accessed after its complete cohort training stage; this older test is not the new formal target stream for any of the three RCI reductions.

\begin{table}[htbp]\centering\small
\caption{Historical checkpoint selection, verified from saved checkpoint metadata and validation histories. These errors are validation quantities from original training, not new RCI test results. All five runs completed 24,000 updates.}\label{tab:original-checkpoints}
\begin{tabular}{lrr}\toprule
Seed & Selected update & Historical validation error\\\midrule
1000 & 23,900 & 0.116847\\
1001 & 22,900 & 0.075730\\
1002 & 24,000 & 0.109782\\
1003 & 23,600 & 0.111996\\
1004 & 22,900 & 0.080101\\
\bottomrule\end{tabular}
\end{table}

\paragraph{Verification and unavailable information.}
Checkpoint, configuration, implementation, and driver hashes match the original archives. Optimizer fields corroborate Adam; all 241 history entries per seed agree with the validation schedule and selected minimum. Source, configuration, protocol, manifest, histories, checkpoints, and audit are under \texttt{evidence/training/}.

The full historical sample stream and exact original Python, PyTorch, and NumPy versions are unavailable in the inspected records; current versions are not substituted. We cannot claim bitwise retraining across platforms or complete historical target nonexposure. These gaps do not alter checkpoint identity. No training or new model evaluation was used in this revision.

\section{Exploratory horizon diagnostics from archived trajectories}\label{app:horizon}
The saved v3 trajectories are reaggregated descriptively with all original targets, repeats, models, and maps. No horizon is selected to rescue a gate; the 32-step contract and its inference remain unchanged.

For a stored horizon $\tau\in\{1,\ldots,32\}$, define
\begin{equation}
 C_a(\tau)=\max_{t<\tau}\max_j\|r_{a,j}(t)\|_2/(2\sqrt8),\qquad
 D_a(\tau)=\|q(X_\tau^a)-\overline q(Z_\tau^a)\|_2/\sqrt2.
\end{equation}
The displayed curves are means over the fixed five models and the original target/repeat distribution. They are not predictions averaged before scoring. Gains at each time use the original normalized task loss with the same target. All 33 saved output times and all 32 saved residual steps are included in the accompanying CSV, including individual-model summaries. At time zero the empty prefix-maximum diagnostic is assigned zero. Table~\ref{tab:horizon} displays four checkpoints; the time-32 means reproduce the original primary statistics within $2\times10^{-7}$.

\begin{table}[htbp]\centering\scriptsize
\caption{Exploratory saved-trajectory means for v3. These are point estimates, without new certification or simultaneous inference across horizons.}\label{tab:horizon}
\begin{tabular}{rrrrrrr}\toprule
$\tau$ & $\widehat C_1(\tau)$ & $\widehat C_0(\tau)$ & $\widehat D_1(\tau)$ & $\widehat D_0(\tau)$ & $\widehat G_{\rm micro}(\tau)$ & $\widehat G_{\rm macro}(\tau)$\\\midrule
8 & 0.055110 & 0.059756 & 0.017473 & 0.016744 & 0.020920 & 0.014443\\
16 & 0.057891 & 0.061872 & 0.036579 & 0.061577 & 0.076409 & 0.059628\\
24 & 0.059395 & 0.062639 & 0.068769 & 0.090185 & 0.184230 & 0.142408\\
32 & 0.059428 & 0.062696 & 0.085890 & 0.082877 & 0.257976 & 0.217890\\
\bottomrule\end{tabular}
\end{table}

\begin{figure}[htbp]\centering
\IfFileExists{figures/rci_horizon_diagnostics.png}{\includegraphics[width=.94\linewidth]{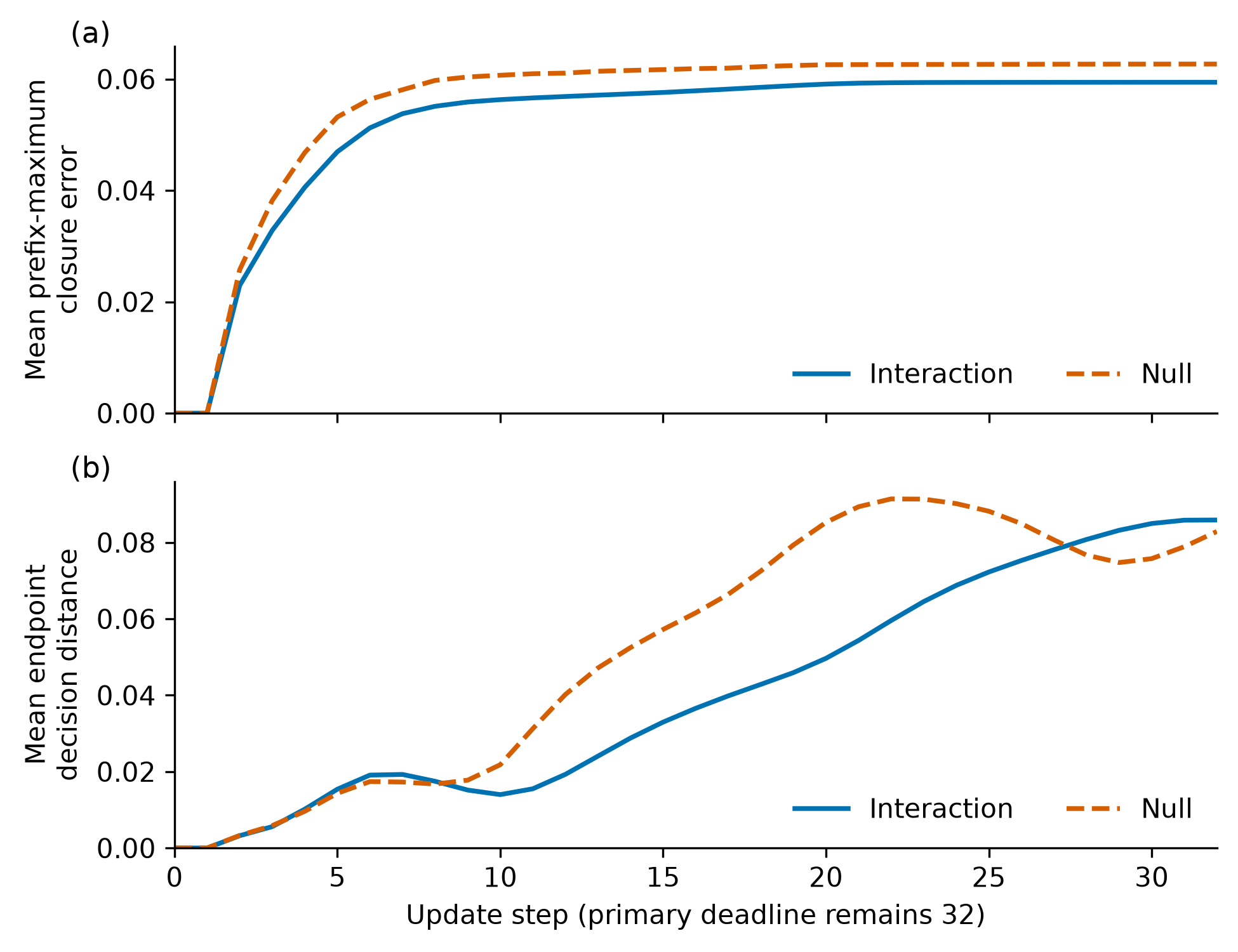}}{\fbox{\parbox{.88\linewidth}{\centering Figure supplied with the source package: \texttt{figures/rci\_horizon\_diagnostics.png}. Numerical results are included in the tables and text.}}}
\caption{Descriptive diagnostics from unchanged archived v3 paths: (a) mean prefix-maximum closure defect; (b) mean endpoint decision discrepancy. Both arms and all five models are retained. Panel (a) is nondecreasing by definition of a prefix maximum; this is not evidence of linear accumulation or dynamical instability. Panel (b) need not be monotone. No horizon is selected from these curves, and $T=32$ remains the primary deadline.}\label{fig:horizon}
\end{figure}

Table~\ref{tab:horizon} shows that closure maxima are near their final values by time 16, while endpoint discrepancies continue to change. Null discrepancy falls from time 24 to 32, excluding a monotone-error interpretation. Shorter deadlines also have smaller task gains, so smaller early discrepancies alone do not certify mature capability preservation. These diagnostics establish neither a causal mechanism nor long-horizon stability.

The evidence index links \texttt{evidence/horizon\_reanalysis/horizon\_metrics.csv} and its source manifest, which records hashes of the five original saved-array files and the target array. The manifest explicitly separates this descriptive reaggregation from the frozen primary result.

\end{document}